\documentclass[11pt]{article}

\usepackage[english]{babel}

\usepackage[a4paper,top=2cm,bottom=2cm,left=3cm,right=3cm,marginparwidth=1.75cm]{geometry}

\usepackage{amsmath}
\usepackage{graphicx}
\usepackage[colorlinks=true, allcolors=blue]{hyperref}

\usepackage{graphicx}
\usepackage{tcolorbox}
\usepackage{graphicx}
\usepackage{empheq}

\usepackage{tikz}
\usetikzlibrary{math} 
\usepackage{import}
\usepackage{float}
\usepackage{placeins} 
\usepackage{xcolor}
\usepackage{amsmath}
\usepackage{multirow}
\usepackage{amsmath,amsfonts,amssymb,amsthm}
\usepackage{mathtools}
\usepackage[linesnumbered,ruled]{algorithm2e}
\usepackage{mathtools}
\usepackage[super]{nth}
\usepackage[font=footnotesize,labelfont=bf]{caption}
\usepackage{subcaption}
\usepackage{graphicx}
\usepackage{wrapfig}
\usepackage{verbatim} 
\usepackage{longtable}
\usepackage{array}
\usepackage{supertabular}
\usepackage{longtable}
\usepackage{tabularx}
\usepackage{multirow}
\usepackage{colortbl}
\usepackage{fancyhdr}
\usepackage{xcolor}
\usepackage{pgf,tikz}
\usepackage{pgfplots, pgfplotstable}

\usepackage{multido}
\usepackage{calc}
\usepackage{fp}
\usepackage{xifthen}
\usepackage{xargs}
\usepackage{etoolbox}
\DeclareMathAlphabet{\mathpzc}{OT1}{pzc}{m}{it}
\usepackage{breakcites} 
\usepackage[utf8]{inputenc}
\usepackage{helvet}
\usepackage{geometry}
\usepackage{amssymb}
\usepackage{enumitem}
\usepackage{tabularx}
\usepackage{xcolor}
\usepackage[absolute,overlay]{textpos}
\usepackage{graphicx}
\usepackage{lipsum}
\usepackage{caption}
\usepackage{multicol}
\usepackage{afterpage}
\usepackage{setspace}
\usepackage{pgffor}
\usepackage{parskip}
\usepackage{xfrac,nicefrac}
\usepackage{booktabs}

\usepackage{authblk}

\pgfplotsset{compat=1.18}

\usepackage[style=numeric-comp, sorting=none, giveninits=true, maxnames=99]{biblatex}
\renewbibmacro{in:}{} 
\title{FFT-based surrogate modeling of auxetic metamaterials with real-time prediction of effective elastic properties and swift inverse design}

\begin{document}
\author[1]{ Hooman Danesh\thanks {hooman.danesh@rwth-aachen.de}}
\author[2,3]{Daniele Di Lorenzo}
\author[2,3,4]{Francisco Chinesta}
\author[1,5]{Stefanie Reese}
\author[1]{Tim Brepols}
\date{} 

\affil[1]{\footnotesize{Institute of Applied Mechanics, RWTH Aachen University, Mies-van-der-Rohe-Str. 1, 52074 Aachen, Germany}}
\affil[2]{\footnotesize{ESI Group, Symbiose 2, 10 Av. Aristide Briand, 92220 Bagneux, France}}
\affil[3]{\footnotesize{
PIMM Lab, ENSAM Institute of Technology, 151 Boulevard de l’Hôpital, 75013 Paris, France}}
\affil[4]{\footnotesize{CNRS@CREATE,1 CREATE Way, 04-05 CREATE Tower, Singapore 138602}}
\affil[5]{\footnotesize{University of Siegen, Adolf-Reichwein-Str. 2a, 57076 Siegen, Germany}}

\maketitle

\begin{abstract}
Auxetic structures, known for their negative Poisson's ratio, exhibit effective elastic properties heavily influenced by their underlying structural geometry and base material properties. While periodic homogenization of auxetic unit cells can be used to investigate these properties, it is computationally expensive and limits design space exploration and inverse analysis. In this paper, surrogate models are developed for the real-time prediction of the effective elastic properties of auxetic unit cells with orthogonal voids of different shapes. The unit cells feature orthogonal voids in four distinct shapes, including rectangular, diamond, oval, and peanut-shaped voids, each characterized by specific void diameters. The generated surrogate models accept geometric parameters and the elastic properties of the base material as inputs to predict the effective elastic constants in real-time. This rapid evaluation enables a practical inverse analysis framework for obtaining the optimal design parameters that yield the desired effective response. The fast Fourier transform (FFT)-based homogenization approach is adopted to efficiently generate data for developing the surrogate models, bypassing concerns about periodic mesh generation and boundary conditions typically associated with the finite element method (FEM). The performance of the generated surrogate models is rigorously examined through a train/test split methodology, a parametric study, and an inverse problem. Finally, a graphical user interface (GUI) is developed, offering real-time prediction of the effective tangent stiffness and performing inverse analysis to determine optimal geometric parameters.\\
\\
\textbf{Keywords:} 
Auxetic structures, Surrogate models, Effective properties, Inverse analysis, FFT-based homogenization

\end{abstract}

\section{Introduction}
\label{sec:1}

\subsection{Context and rationale}
\label{sec:1.1}

Auxetic materials represent a fascinating frontier in materials science and engineering, distinguished by their unique and counter-intuitive property of exhibiting negative Poisson's ratios. Unlike conventional materials that contract laterally when stretched, auxetic materials exhibit the remarkable property of expanding in the direction transverse to the applied tension. This counter-intuitive feature has a wide range of innovative applications, including energy absorption \cite{zhang2020large,najafi2021experimental,choudhry2022plane}, flexible electronics \cite{yan2019novel,jiang2022flexible,jang2022auxetic}, protective sports equipment \cite{allen2015auxetic,duncan2018review,tahir2022auxetic}, and medical implants \cite{kolken2018rationally,kolken2020mechanical,shirzad2023auxetic}. Such an extraordinary behavior, i.e., negative Poisson's ratio, stems from the unique structural geometry of auxetic materials. The effective response of the designed auxetic structures is substantially impacted by changing the base material properties and the unit cell geometry. Each engineering application demands a unique set of auxetic structures that possess certain effective features. For this reason, it would be desirable to establish a framework that can extract the effective properties of auxetics with varying geometries and base constituents.

Extensive research has been conducted on the homogenization and extraction of the effective properties of auxetic metamaterials using various analytical, experimental, and numerical approaches. An analytical model to predict the effective elastic constants and Poisson's ratio of hexagonal and re-entrant honeycomb structures was developed in \cite{masters1996models}. Using a discrete asymptotic homogenization approach, Dos Reis and Ganghoffer \cite{dos2012equivalent} provided closed form expressions for the effective elastic moduli and Poisson's ratio of re-entrant and chiral structures formed from a network of beams. A theoretical model for beam networks of 3D re-entrant honeycomb auxetic structures was developed in \cite{yang2015mechanical}, providing analytical solutions for the elastic modulus, Poisson’s ratio, and yield strength. This model was validated through comparison with experimental and finite element (FE) simulation results. Chan and Evans \cite{chan1999mechanical1,chan1999mechanical2} carried out experimental studies to investigate the elastic response of auxetic polymeric foams under tension, compression, and shear. Combined FE simulations and experimental testing investigations of the mechanical properties of re-entrant cell honeycombs under uniaxial loading were performed in \cite{scarpa2000numerical}. Dirrenberger et al. \cite{dirrenberger2013effective} employed the finite element method (FEM) combined with periodic homogenization to extract the effective elastic moduli of chiral and honeycomb lattices and to examine their anisotropy. The homogenized mechanical properties of hyperelastic inclusion-matrix composites were obtained via numerical homogenization by Kochmann and Venturini \cite{kochmann2013homogenized}. Their study revealed that auxeticity in such finitely strained composites emerged for certain geometries and base material properties. Slann et al. \cite{slann2015cellular} developed a numerical asymptotic homogenization FE approach to conduct a parametric study on the mechanical properties of rectangular void auxetic unit cells. The model's accuracy was verified by comparing its predictions with experimental results obtained from uniaxial tensile tests on perforated sheets. Numerical and experimental investigations of auxetic structures with randomly oriented perforations were performed by Grima et al. \cite{grima2016auxetic}.  They employed the FE approach and periodic boundary conditions to extract the average response of these auxetics.

With regard to the previously mentioned analytical, numerical, and experimental methods used to study the effective response of auxetic metamaterials, it is crucial to mention their limitations in specific situations. Analytical approaches are mainly limited to the extraction of the effective properties for simple discrete structures, such as those formed from a network of  beams. Furthermore, when it comes to design space exploration, real-time prediction of effective properties, or performing inverse design, numerical simulations or experimental studies present significant challenges. Numerical simulations are computationally expensive, especially for inverse problems, while experimental methods are impractical and often too costly for iterative design processes. Consequently, there is a critical need for surrogate models that can provide real-time evaluations of effective properties for auxetic structures. These models, which are trained on high-fidelity datasets, enable rapid design space exploration, sensitivity analyses, and inverse design of auxetic structures.

Random forests regression was used by Tajalasir et al. \cite{tajalsir2022numerical} to develop machine learning models for predicting the dynamic impact stress of hierarchical auxetic structures. In a related study, Ben-Yelun et al. \cite{ben2023gam} employed the random forests method to create surrogate models for estimating the effective elastic modulus and Poisson's ratio of a class of tunable 3D auxetic metamaterials, using geometric parameters as inputs. Neural networks were used in \cite{lyngdoh2022elucidating} to develop machine learning models, trained on large datasets produced by FE simulations, to predict Poisson's ratio of cementitious oval void axuetics. In another study \cite{vyavahare2023fdm}, machine learning models were developed based on experimental data to predict the strength, stiffness, and energy absorption of re-entrant auxetic structures manufactured by fused deposition modeling. A machine learning model to predict the effective Poisson's ratio of oval void auxetic structures was developed in \cite{wang2021novel}. The accuracy of the model was examined against experiments and FE simulations. Du et al. \cite{du2023auxetic} employed recurrent neural networks to predict the effective Poisson's ratio of kirigami auxetics based on four geometric input parameters. A neural-network-based surrogate model, trained on datasets from isogeometric analysis homogenization, was developed in \cite{liao2022deep} to establish an efficient inverse design framework for designing tetra-chiral auxetics with target Poisson's ratios. Chang et al. \cite{chang2022machine} employed back-propagation neural networks and the genetic algorithm to develop a model for the inverse design of re-entrant auxetics with zero Poisson's ratio. Similarly, the same machine learning algorithms were used in \cite{liu2023high} to set up an efficient data-driven framework for the optimization design of auxetic structures with peanut-shaped perforations. Machine-learning-based surrogate models for auxetic structures have been extensively developed in recent literature, which are not within the scope of this paper. For a comprehensive overview of these advancements, interested readers are referred to \cite{zhang2024critical}.

\subsection{In silico data generation}

Efficient data generation is crucial in the development of accurate and robust surrogate models, including those for auxetic metamaterials. While FEM is commonly employed for generating synthetic data in the development of surrogate models, an alternative approach exists that has been largely overlooked in this context. The fast Fourier transform (FFT)-based method offers a highly efficient and rapid means of data generation, presenting significant potential for enhancing the process of building surrogate models for auxetic metamaterials. Owing to their matrix-free formulation, FFT-based methods can be pretty memory-efficient, allowing for rather large systems to be homogenized in comparison to FEM. In FFT-based homogenization, periodicity follows naturally from the definition of the Fourier transform and does not need extra care, as in FEM. This intrinsic periodicity eliminates the need to check for symmetric node placement on boundaries, which is crucial in FEM for applying periodic boundary conditions. 

The FFT-based homogenization dates back to the primary works of Moulinec and Suquet \cite{moulinec1994fast,moulinec1998numerical} on the effective mechanical response of composites. The FFT-based scheme is formulated in a convolution integral form, which is known as the Lippmann-Schwinger equation, and is then transformed to a more manageable algebraic product form in Fourier space (see, e.g., \cite{schneider2021review,lucarini2021fft} for review). In spite of its well-known advantages, the basic FFT scheme of Moulinec and Suquet \cite{moulinec1994fast,moulinec1998numerical} reaches a converged solution notably dependent on the difference in stiffness between the phases found in the microstructure. This means that the method converges faster when the stiffness contrast between the phases is smaller. In the context of metamaterial unit cells, including the auxetic structures in the present study, one of these phases symbolizes an empty space possessing zero stiffness. Consequently, an infinite stiffness contrast arises, impeding the convergence of the basic scheme when applied to such materials \cite{danesh2023challenges,lucarini2022adaptation}. To address this limitation, Lucarini et al. \cite{lucarini2022adaptation} proposed a solution involving the utilization of the Galerkin FFT homogenization method \cite{vondvrejc2014fft,vondvrejc2015guaranteed}, along with the rotated finite difference grid \cite{willot2015fourier} and the minimal residual (MINRES) solver. Such an approach is employed in the present work to generate large high-fidelity datasets by solving the homogenization problem for different auxetic structures.

\subsection{Objectives and contributions}
\label{sec:1.2}

Although FFT-based homogenization approaches are relatively straightforward to implement and use, the computational cost becomes prohibitive when it comes to real-time parametric studies and inverse design. Within this context, parametric studies enable engineers to rapidly evaluate how changes in geometric and base material parameters affect the effective elastic properties of auxetic structures, facilitating quick sensitivity analyses and understanding parameter-property relationships. Equally important is inverse design, where desired material properties, such as target values for the effective Poisson's ratio or other elastic constants, are defined, and the optimal structures meeting these properties are found using optimization algorithms. All these tasks require numerous evaluations of the homogenization problem for different microstructures. Therefore, the development of fast and accurate surrogate models for real-time prediction of effective properties is crucial. These surrogates enable efficient design space exploration, rapid parametric studies, evaluation of potential unit cell configurations, and swift solution of inverse problems without resorting to time-consuming and computationally expensive simulations.

In the present work, surrogate models are developed for four types of orthogonal void auxetic unit cells with different perforations, including rectangular, diamond, oval, and peanut-shaped voids. The efficient FFT-based homogenization method is used to generate large high-fidelity datasets required to build the surrogate models. This approach represents a novel and advantageous method for data generation in this context. The exceptional efficiency of this method allows for the creation of substantial datasets within a relatively short period of time. Consequently, the accuracy of the surrogate models can be significantly enhanced due to the voluminous data provided, marking a notable improvement over traditional data generation techniques in this field. The random forests regression method is used to develop reasonably accurate surrogates, accepting the geometric and base material parameters as input and predicting the homogenized effective tangent stiffness in real time. This real-time prediction capability facilitates effective probe of design spaces, allowing for rapid parametric studies and sensitivity analyses. An inverse problem framework is also established using a straightforward brute-force algorithm. While many previous studies have primarily focused on the effective Poisson's ratio as the target value for the inverse problem, the current approach offers the flexibility to select any component or any combination of the components of the effective elastic stiffness as the target output. The performance of the models is examined through a train/test split approach, a parametric study, and an inverse problem. It is demonstrated that the generated surrogate models can provide highly accurate predictions in close agreement with the actual values obtained from the FFT-based solver. The capability of the established framework in swiftly solving the inverse problem is shown for the rectangular void unit cell. Furthermore, this study also presents a graphical user interface (GUI) to enhance the accessibility and practical application of the developed surrogate models. This interface enables the real-time prediction of effective elastic constants and facilitates the inverse design of auxetic metamaterials, providing an efficient tool for both analyses and design processes with visual feedback.

The remainder of the present paper is structured as follows: Section \ref{sec:2} presents the theoretical basis of the FFT-based homogenization and, specifically, the computation of the homogenized effective tangent stiffness. Section \ref{sec:3} is concerned with the surrogate modeling of the given auxetic structures, introducing the input parameters and output quantities of the models as well as the random forests regression method. The inverse analysis framework, including the brute-force algorithm, is explained in Section \ref{sec:4}. Two case studies are presented in Section \ref{sec:5}, including a parametric study to assess the accuracy of the generated models and to investigate the auxetic behavior of various unit cell geometries, followed by a practical inverse problem to achieve specific mechanical properties. Section \ref{sec:6} introduces the developed GUI, highlighting its user-friendly capabilities in real-time prediction of effective stiffness and inverse analyses. Finally, Section \ref{sec:7} summarizes concluding remarks and future research directions.

\section{FFT-based homogenization}
\label{sec:2}

Following \cite{vondvrejc2014fft,vondvrejc2015guaranteed,lucarini2022adaptation}, the initial step in the Galerkin FFT method is to apply the principle of virtual work to formulate the weak form of the boundary value problem. Considering small strain kinematics, the weak form, in the absence of boundary traction terms due to periodic boundary conditions, is written as

\begin{equation}
\label{eq:1}
    \int_{\Omega} \delta \boldsymbol{\varepsilon}(\boldsymbol{x}): \boldsymbol{\sigma}(\boldsymbol{x}) \, \mathrm{d} \Omega=0,
\end{equation}

where $\Omega$ represents the microscopic domain, and $\boldsymbol{\sigma}(\boldsymbol{x})$ is the microscopic stress tensor at the microscopic position $\boldsymbol{x}$. The microscopic virtual strain $\delta \boldsymbol{\varepsilon}(\boldsymbol{x})$ needs to satisfy the periodicity, symmetry, and compatibility conditions. The last two conditions can be applied by employing a projection operator $\mathbb{G}$ (for more details, cf. \cite{vondvrejc2014fft}) as

\begin{equation}
\label{eq:2}
    \delta \boldsymbol{\varepsilon}(\boldsymbol{x})=(\mathbb{G} * \boldsymbol{\zeta})(\boldsymbol{x}).
\end{equation}

In the above, $\boldsymbol{\zeta}(\boldsymbol{x})$ is an arbitrary second-order tensor field, whose symmetry and compatibility are imposed by the fourth-order projection tensor $\mathbb{G}$ through the convolution $*$ in real space. By inserting Eq. \ref{eq:2} into Eq. \ref{eq:1} and considering the major symmetry of $\mathbb{G}$ (i.e., $G_{ijkl}=G_{klij}$), the following integral equation emerges:

\begin{equation}
\label{eq:3}
    \int_{\Omega} \boldsymbol{\zeta}(\boldsymbol{x}) : \left[ (\mathbb{G} * \boldsymbol{\sigma})(\boldsymbol{x}) \right] \, \mathrm{d} \Omega=0.
\end{equation}

As Eq. \ref{eq:3} needs to be fulfilled for any arbitrary test function $\boldsymbol{\zeta}(\boldsymbol{x})$, the equilibrium in its weak form reduces to the following equation, which is valid at every point in domain $\Omega$:

\begin{equation}
\label{eq:4}
    \mathbb{G} * \boldsymbol{\sigma} = \mathbf{0}.
\end{equation}

Performing the forward ($\mathcal{F}$) and backward ($\mathcal{F}^{-1}$) Fourier transforms and taking the convolution property of Fourier transform \cite{lucarini2021fft} into account, Eq. \ref{eq:4} can be written as

\begin{equation}
\label{eq:5}
    \mathbb{G} * \boldsymbol{\sigma}=\mathcal{F}^{-1} \left[ \widehat{\mathbb{G}}: \mathcal{F}(\boldsymbol{\sigma}) \right] = \mathbf{0},
\end{equation}

where $\widehat{\mathbb{G}}$ is the projection tensor in the Fourier space, whose components are expressed as follows \cite{vondvrejc2014fft,vondvrejc2015guaranteed,lucarini2022adaptation}:

\begin{equation}
\label{eq:6}
    \hat{G}_{i j k l}=\left\{\begin{array}{lr}
    0, & \text { for zero and Nyquist frequencies } \\[1em]
    {\left[I_{i p k q}^s \xi_p \xi_q\right]^{-1} \xi_j \xi_{l,}} & \text { for other frequencies }
    \end{array}\right.
\end{equation}

In Eq. \ref{eq:6}, $\mathbb{I}^s$ represents the fourth-order symmetric identity tensor, and $\boldsymbol{\xi}$ is the frequency vector, which can be written as

\begin{equation}
\label{eq:7}
    \xi_m=\mathrm{i} q_m N_m / L_m, \quad \text { where } \quad q_m=2 \pi\left(n_m-N_m / 2\right) / N_m \quad \text { with } \quad n_m=0,1, \ldots, N_m-1.
\end{equation}

Here, $m=1,2$ and $m=1,2,3$ for 2D and 3D domains, respectively, $\mathrm{i}$ is reserved for the unit imaginary number, $L_m$ represents the length of the unit cell edge in the $m$ direction, and $N_m$ expresses the number of voxels/pixels in this direction.

As suggested in \cite{lucarini2022adaptation}, in order to reach better convergence for materials with high stiffness contrast (e.g., the auxetic structures in the current study), Willot’s rotated finite difference discretization \cite{willot2015fourier} is used. In this case, the spatial frequency vector $\boldsymbol{\xi}$ is modified as

\begin{equation}
\label{eq:8}
    \xi_m^{\prime}=\mathrm{i} 2 N_m / L_m \tan \left(q_m / 2\right) \prod_{m=1}^d \frac{1}{2}\left(1+e^{-\mathrm{i} q_m}\right), \quad \text { where } \quad q_m=2 \pi\left(n_m-N_m / 2\right) / N_m.
\end{equation}

Next, assuming linear elastic constitutive behavior and using the additive decomposition of the microscopic strain field $\boldsymbol{\varepsilon}(\boldsymbol{x})$ into macroscopic strain $\bar{\boldsymbol{\varepsilon}}$ and fluctuation strain $\tilde{\boldsymbol{\varepsilon}}(\boldsymbol{x})$, the stress-strain constitutive relation can be written as

\begin{equation}
\label{eq:9}
    \boldsymbol{\sigma}(\boldsymbol{x}) = \mathbb{D} : \left[ \bar{\boldsymbol{\varepsilon}} + \tilde{\boldsymbol{\varepsilon}}(\boldsymbol{x}) \right],
\end{equation}

where $\mathbb{D}$ is the fourth-order elastic stiffness tensor, and the macroscopic strain field $\bar{\boldsymbol{\varepsilon}}$ is defined by volume averaging through

\begin{equation}
\label{eq:10}
    \bar{\boldsymbol{\varepsilon}} = \left< \boldsymbol{\varepsilon} \right> = \frac{1}{V} \int_V {\boldsymbol{\varepsilon}}(\boldsymbol{x}) \, \mathrm{d}V.
\end{equation}

Here, $\left< \cdot \right>$ denotes the average over the volume. Substituting Eq. \ref{eq:9} into Eq. \ref{eq:5}, the equilibrium equation can be rewritten in the form

\begin{equation}
\label{eq:11}
    \mathcal{F}^{-1}[\widehat{\mathbb{G}}: \mathcal{F}(\mathbb{D}: \tilde{\boldsymbol{\varepsilon}})]=-\mathcal{F}^{-1}[\widehat{\mathbb{G}}: \mathcal{F}(\mathbb{D}: \bar{\boldsymbol{\varepsilon}})].
\end{equation}

Finally, Eq. \ref{eq:11} can be solved using any Newton-Krylov solver; however, the MINRES solver is employed here for auxetic structures with infinite stiffness contrast \cite{lucarini2022adaptation}. After computing the microscopic stress field $\boldsymbol{\sigma}(\boldsymbol{x})$, the macroscopic average stress $\bar{\boldsymbol{\sigma}}$ is obtained by

\begin{equation}
\label{eq:12}
    \bar{\boldsymbol{\sigma}} = \left< \boldsymbol{\sigma} \right> = \frac{1}{V} \int_V {\boldsymbol{\sigma}}(\boldsymbol{x}) \, \mathrm{d}V.
\end{equation}

\subsection{Homogenized effective tangent stiffness}
The formulations presented in Section \ref{sec:2} can be used to compute the full-field solutions for the microscopic strain field $\boldsymbol{\varepsilon}(\boldsymbol{x})$ and the microscopic stress field $\boldsymbol{\sigma}(\boldsymbol{x})$ as well as their corresponding macroscopic average values $\bar{\boldsymbol{\varepsilon}}$ and $\bar{\boldsymbol{\sigma}}$. However, in the current study, the homogenized effective tangent stiffness $\mathbb{C}$ is the main quantity of interest to compute. This homogenized effective tangent stiffness $\mathbb{C}$ is defined by

\begin{equation}
\label{eq:13}
    \mathbb{C} = \frac{\partial\bar{\boldsymbol{\sigma}}}{\partial\bar{\boldsymbol{\varepsilon}}}.
\end{equation}

When linear elasticity is concerned, a straightforward way to determine the homogenized effective stiffness tensor $\mathbb{C}$ is through a simple perturbation method \cite{idrissi2022multiparametric,gierden2022review}. This involves calculating the macroscopic average stress $\bar{\sigma}_{ij}$ induced by six (in 3D) or three (in 2D) macroscopic strain loadings $\bar{\varepsilon}_{kl} = \beta$. Here, $\beta$ is a real constant, which can be simply set to $\beta=1$ since its value is irrelevant within linear elasticity. The unperturbed state is considered to be the state of zero strain. Consequently, the components of the homogenized effective tangent stiffness $C_{ijkl}$ can be computed by

\begin{equation}
\label{eq:14}
    C_{ijkl} = \frac{\bar{\sigma}_{ij(kl)}}{\beta},
\end{equation}

where $\bar{\sigma}_{ij(kl)}$ refers to the macroscopic stress component $ij$ resulting from the application of a perturbation in the macroscopic strain component $kl$ while all other strain components remain zero.

An alternative approach to computing the homogenized effective stiffness tensor $\mathbb{C}$ is to directly solve for it using the projection operator $\mathbb{G}$ \cite{gokuzum2018algorithmically,rambausek2019two,minh2020surrogate}.  To this end, keeping Eqs. \ref{eq:10} and \ref{eq:12} in mind and taking advantage of the additive split of strain, i.e., $\boldsymbol{\varepsilon}(\boldsymbol{x}) = \bar{\boldsymbol{\varepsilon}} + \tilde{\boldsymbol{\varepsilon}}(\boldsymbol{x})$,  Eq. \ref{eq:13} can be rewritten as

\begin{equation}
\label{eq:15}
    \mathbb{C} = \left< \frac{\partial\boldsymbol{\sigma}}{\partial\boldsymbol{\varepsilon}} : \left( \mathbb{I} + \frac{\partial\tilde{\boldsymbol{\varepsilon}}}{\partial\bar{\boldsymbol{\varepsilon}}}\right) \right> = \frac{1}{V} \int_V \left[ \frac{\partial\boldsymbol{\sigma}}{\partial\boldsymbol{\varepsilon}} : \left( \mathbb{I} + \frac{\partial\tilde{\boldsymbol{\varepsilon}}}{\partial\bar{\boldsymbol{\varepsilon}}}\right) \right] \, \mathrm{d}V.
\end{equation}

Given that ${\partial\boldsymbol{\sigma}}/{\partial\boldsymbol{\varepsilon}} = \mathbb{D}$, the only term in Eq. \ref{eq:15} that needs to be computed to achieve the homogenized effective stiffness $\mathbb{C}$ is the sensitivity of the strain fluctuation field with respect to the strain average field, i.e., ${\partial\tilde{\boldsymbol{\varepsilon}}}/{\partial\bar{\boldsymbol{\varepsilon}}}$. This term will appear by substituting Eq. \ref{eq:9} into Eq. \ref{eq:4} and, subsequently, differentiating with respect to the macroscopic strain field $\bar{\boldsymbol{\varepsilon}}$:

\begin{equation}
\label{eq:16}
    \mathbb{G} *\left[\mathbb{D}:\left(\mathbb{I}+\frac{\partial \tilde{\boldsymbol{\varepsilon}}}{\partial \bar{\boldsymbol{\varepsilon}}}\right)\right]=\mathbf{0}.
\end{equation}

Similar to the equilibrium equation, i.e., Eqs. \ref{eq:4}, \ref{eq:5} and \ref{eq:11}, using the forward and inverse Fourier transforms, Eq. \ref{eq:16} can also be rewritten as 

\begin{equation}
\label{eq:17}
    \mathcal{F}^{-1}\left[ \widehat{\mathbb{G}}: \mathcal{F}\left( \mathbb{D} : \frac{\partial \tilde{\boldsymbol{\varepsilon}}}{\partial \bar{\boldsymbol{\varepsilon}}}\right) \right]=-\mathcal{F}^{-1}\left[ \widehat{\mathbb{G}}: \mathcal{F}\left( \mathbb{D}\right) \right].
\end{equation}

Eventually, analogous to the solution method employed for Eq. \ref{eq:11}, Eq. \ref{eq:17} is also solved using the MINRES Newton-Krylov solver, which is particularly effective for auxetic structures with infinite stiffness contrast. In the present study, this second approach, i.e., adopting the projection operator $\mathbb{G}$ and Eqs. \ref{eq:15} and \ref{eq:17}, is used to compute the homogenized effective tangent stiffness $\mathbb{C}$.

\section{Surrogate modeling}
\label{sec:3}

\begin{figure}[h!]  
    \centering
    \begin{subfigure}{0.22\textwidth}  
        \centering  
        \includegraphics[width=0.8\textwidth]{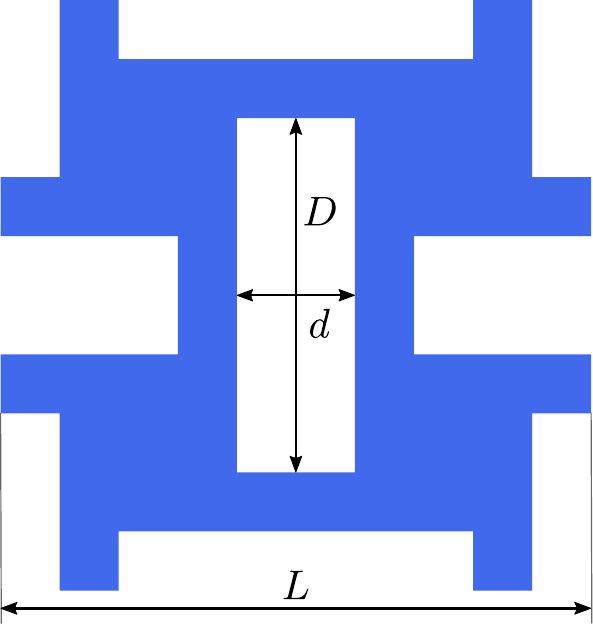}  
        \caption{Rectangular void}  
        \label{fig:1a}  
    \end{subfigure}  
    \begin{subfigure}{0.22\textwidth}  
        \centering  
        \includegraphics[width=0.8\textwidth]{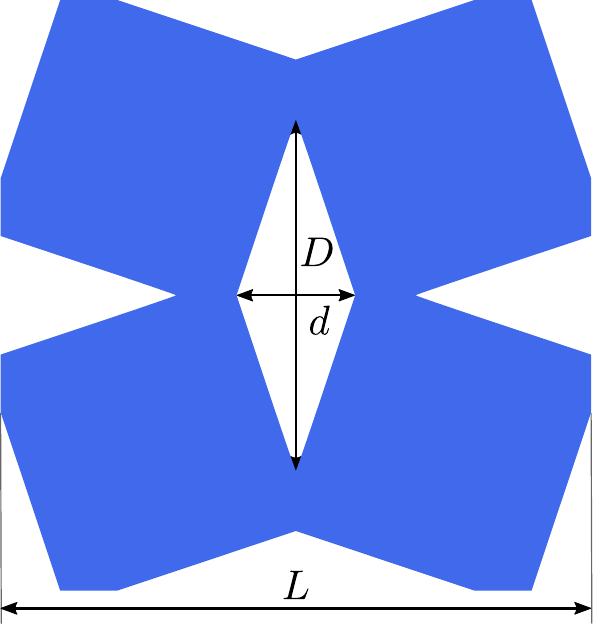}  
        \caption{Diamond void}  
        \label{fig:1b}  
    \end{subfigure}
    \begin{subfigure}{0.22\textwidth}  
        \centering  
        \includegraphics[width=0.8\textwidth]{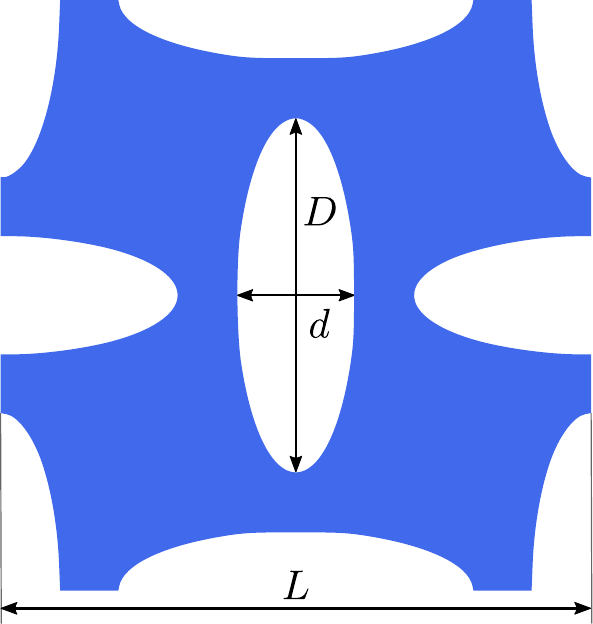}  
        \caption{Oval void}  
        \label{fig:1c}  
    \end{subfigure}
    \begin{subfigure}{0.22\textwidth}  
        \centering  
        \includegraphics[width=0.8\textwidth]{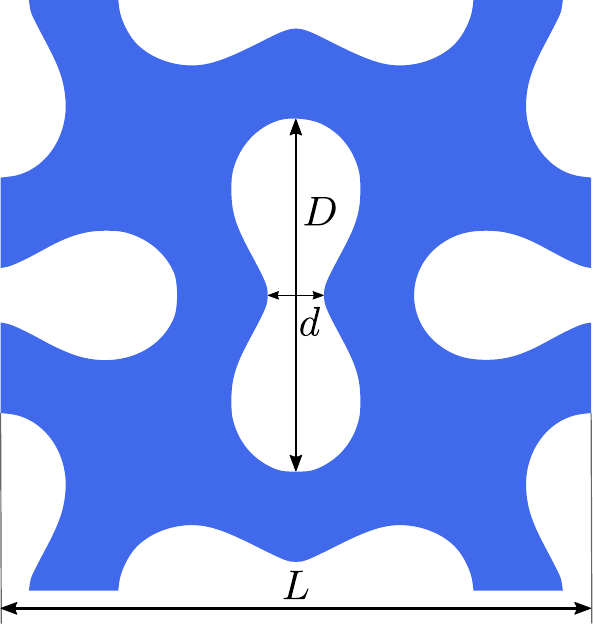}  
        \caption{Peanut-shaped void}  
        \label{fig:1d}  
    \end{subfigure}  
    \caption{Auxetic unit cells with orthogonal voids of different shapes, characterized by the unit cell length $L$ and void diameters $d$ and $D$.}
    \label{fig:1}  
\end{figure}

This section is devoted to the surrogate modeling of four different auxetic structures shown in Figure \ref{fig:1}. These auxetic structures are formed from orthogonal voids of different shapes, including rectangular, diamond, oval, and peanut-shaped voids. The input parameters and the output quantities for the desired surrogate models are discussed in the following sections.

\subsection{Input parameters}
\label{sec:3.1}

As shown in Figure \ref{fig:1}, the geometric input parameters for each of the four structures are the relative void diameters $d/L$ and $D/L$. The geometric descriptions of the rectangular, diamond, and oval void unit cells are relatively straightforward. However, the peanut-shaped void requires a more complex mathematical representation. For this particular void geometry, the following mathematical expression in the polar coordinate system is used \cite{reutskiy2012method}:

\begin{equation}
    \rho(\theta)=\sqrt{\cos (2 \theta)+\sqrt{1.1-\sin ^2(2 \theta)}},
\end{equation}

which can be transferred to the Cartesian coordinate system and scaled with the relative void diameters $d/L$ and $D/L$ as

\begin{equation}
x_1=\frac{d}{L} \frac{1}{\sqrt{1+\sqrt{1.1}}}\rho(\theta) \cos \theta, \qquad x_2=\frac{D}{L} \frac{1}{\sqrt{1+\sqrt{1.1}}}\rho(\theta) \sin \theta, \qquad 0 \leq \theta \leq 2 \pi.
\end{equation}

In the above, $\rho(\theta)$ represents the radial distance from the origin to a point on the peanut-shaped curve as a function of the polar angle $\theta$. The variables $x_1$ and $x_2$ denote the Cartesian coordinates of points on the scaled peanut-shaped void. Moreover, the factor $\frac{1}{\sqrt{1+\sqrt{1.1}}}$ is a normalization constant to ensure proper scaling.

Regarding the material input parameters, by examining Hooke's law, it is evident that the microscopic elastic tensor $\mathbb{D}$ is linear in terms of Young's modulus $E$ and nonlinear in terms of Poisson's ratio $\nu$. Consequently, Young's modulus functions as a scaling parameter. This property allows for the simplification of the surrogate models by considering only the effect of Poisson's ratio $\nu$ as the material input parameter, as the effect of Young's modulus can be accounted for through simple scaling of the results subsequent to the prediction.

\subsection{Output quantities}

Considering linear elastic behavior, the constitutive relationship between the macroscopic average stress $\bar{\boldsymbol{\sigma}}$ and the macroscopic average strain $\bar{\boldsymbol{\varepsilon}}$, i.e.,

\begin{equation}
    \bar{\boldsymbol{\sigma}} = \mathbb{C} : \bar{\boldsymbol{\varepsilon}}
\end{equation}

can be written in its most general anisotropic form by using Voigt's notation for the 2D case as

\begin{equation}
    \begin{pmatrix}
        \bar\sigma_{11}\\
        \bar\sigma_{22}\\
        \bar\sigma_{12}
    \end{pmatrix} = 
    \begin{pmatrix}
        C_{11} & C_{12} & C_{13}\\
        C_{21} & C_{22} & C_{23}\\
        C_{31} & C_{32} & C_{33}
    \end{pmatrix}
    \begin{pmatrix}
        \bar\varepsilon_{11}\\
        \bar\varepsilon_{22}\\
        2\bar\varepsilon_{12}
    \end{pmatrix}
\end{equation}

Here, $C_{ij}$ denotes the components of the homogenized effective tangent stiffness $\mathbb{C}$. Analyzing the auxetic structures considered in this study, it is realized that for all four structures, the homogenized effective tangent stiffness exhibits the following form within the stress-strain relationship:

\begin{equation}
\label{eq:22}
    \begin{pmatrix}
        \bar\sigma_{11}\\
        \bar\sigma_{22}\\
        \bar\sigma_{12}
    \end{pmatrix} = 
    \begin{pmatrix}
        C_{11} & C_{12} & 0\\
        C_{12} & C_{11} & 0\\
        0 & 0 & C_{33}
    \end{pmatrix}
    \begin{pmatrix}
        \bar\varepsilon_{11}\\
        \bar\varepsilon_{22}\\
        2\bar\varepsilon_{12}
    \end{pmatrix}
\end{equation}

Such an orthotropic behavior has already been shown in \cite{dirrenberger2013effective} for chiral auxetic structures. It is also to be noted that in contrast to isotropic materials, in which $C_{33}=(C_{11}-C_{12})/2$, such an equality does not hold for the auxetic structures at hand. Therefore, as can be realized from Eq. \ref{eq:22}, the three distinct constants $C_{11}$, $C_{12}$, and $C_{33}$ are required to fully define the homogenized effective tangent stiffness for the given auxetic structures. However, as mentioned in Section \ref{sec:3.1}, due to the linear effect of Young's modulus $E$ on the effective elastic stiffness, the normalized effective elastic constants $C_{11}/E$, $C_{12}/E$, and $C_{33}/E$ are considered as the outputs of the surrogate model. This normalization allows for a more generalized representation of the unit cell's behavior, independent of the specific value of Young's modulus. Besides, the effect of Young's modulus can be incorporated by applying it as a simple scaling factor to the outputs of the generated surrogate models after prediction.

\subsection{Random forests regression}
\label{sec:3.3}

Among others, the random forests regression method \cite{breiman2001random} is used in the current work to generate surrogate models for the four auxetic unit cells presented in Figure \ref{fig:1}. For each unit cell, three separate surrogate models are generated, one for each of the normalized effective elastic constants $C_{11}/E$, $C_{12}/E$, and $C_{33}/E$, resulting in a total of twelve surrogate models. Each model predicts its respective effective elastic constant based on given values of relative void diameters $d/L$ and $D/L$, as well as Poisson's ratio of the base material $\nu$. Random forest regression is an advanced machine learning method that combines multiple decision trees to generate a powerful predictive model. At the training stage, the method creates multiple decision trees, each on a different subset of data, whose predictions are then averaged as the output of the model at the prediction stage. Thus, this makes the random forests approach quite robust in nature by mitigating overfitting of individual decision trees, thereby improving generalization to unseen data.

\begin{figure}[ht]
        \centering
        \begin{subfigure}[t]{1\textwidth}
            \centering
            \includegraphics[width=\linewidth, keepaspectratio]{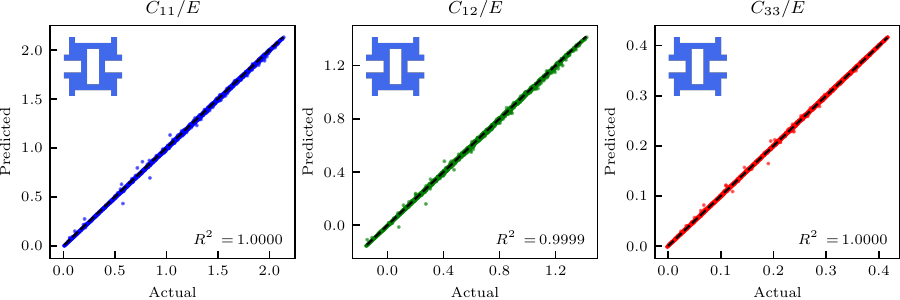}
            \caption{Training dataset}
            \label{fig:2a}
        \end{subfigure}
        \vspace{0.025\textwidth}
        
        \begin{subfigure}[t]{1\textwidth}
            \centering
            \includegraphics[width=\linewidth, keepaspectratio]{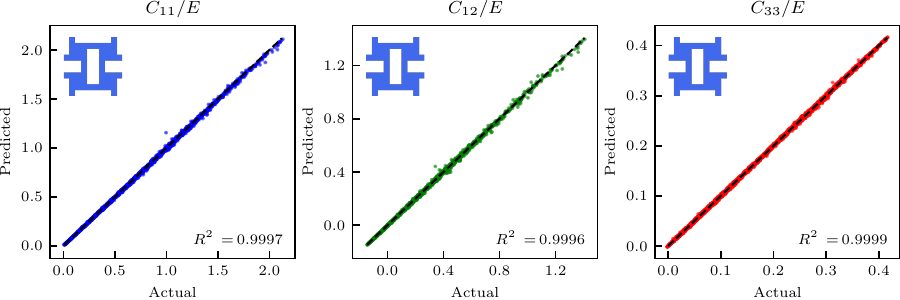}
            \caption{Test dataset}
            \label{fig:2b}
        \end{subfigure}
        \caption{Plots of the predicted values (obtained from the surrogate models) versus the actual values (obtained from the FFT-based solver) of the normalized effective elastic constants $C_{11}/E$, $C_{12}/E$, and $C_{33}/E$ for the rectangular void unit cell. The diagonal dashed line depicts the perfect prediction.}
        \label{fig:2}
\end{figure}

\begin{figure}[ht]
        \centering
        \begin{subfigure}[t]{1\textwidth}
            \centering
            \includegraphics[width=\linewidth, keepaspectratio]{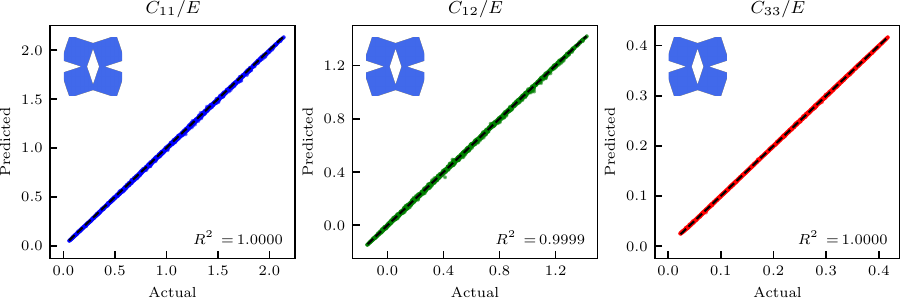}
            \caption{Training dataset}
            \label{fig:3a}
        \end{subfigure}
        \vspace{0.025\textwidth}
        
        \begin{subfigure}[t]{1\textwidth}
            \centering
            \includegraphics[width=\linewidth, keepaspectratio]{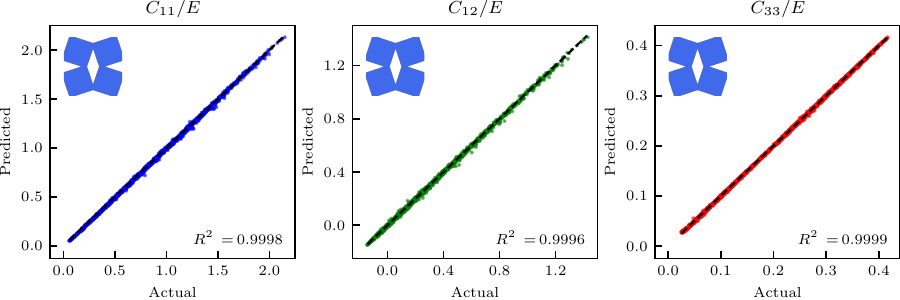}
            \caption{Test dataset}
            \label{fig:3b}
        \end{subfigure}
        \caption{Plots of the predicted values (obtained from the surrogate models) versus the actual values (obtained from the FFT-based solver) of the normalized effective elastic constants $C_{11}/E$, $C_{12}/E$, and $C_{33}/E$ for the diamond void unit cell. The diagonal dashed line depicts the perfect prediction.}
        \label{fig:3}
\end{figure}

\begin{figure}[ht]
        \centering
        \begin{subfigure}[t]{1\textwidth}
            \centering
            \includegraphics[width=\linewidth, keepaspectratio]{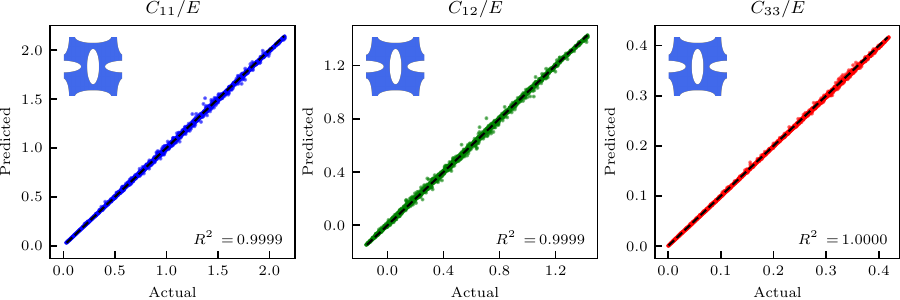}
            \caption{Training dataset}
            \label{fig:4a}
        \end{subfigure}
        \vspace{0.025\textwidth}
        
        \begin{subfigure}[t]{1\textwidth}
            \centering
            \includegraphics[width=\linewidth, keepaspectratio]{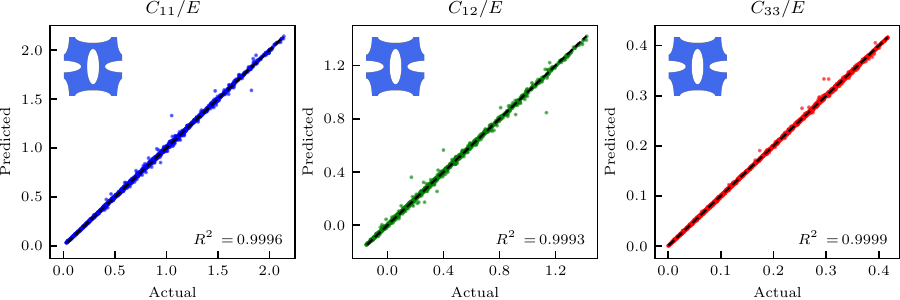}
            \caption{Test dataset}
            \label{fig:4b}
        \end{subfigure}
        \caption{Plots of the predicted values (obtained from the surrogate models) versus the actual values (obtained from the FFT-based solver) of the normalized effective elastic constants $C_{11}/E$, $C_{12}/E$, and $C_{33}/E$ for the oval void unit cell. The diagonal dashed line depicts the perfect prediction.}
        \label{fig:4}
\end{figure}

\begin{figure}[ht]
        \centering
        \begin{subfigure}[t]{1\textwidth}
            \centering
            \includegraphics[width=\linewidth, keepaspectratio]{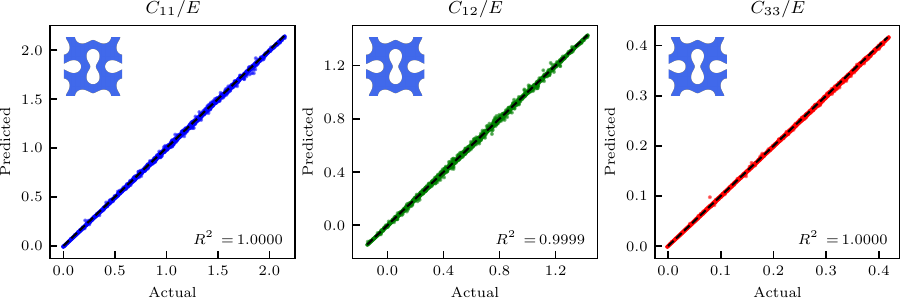}
            \caption{Training dataset}
            \label{fig:5a}
        \end{subfigure}
        \vspace{0.025\textwidth}
        
        \begin{subfigure}[t]{1\textwidth}
            \centering
            \includegraphics[width=\linewidth, keepaspectratio]{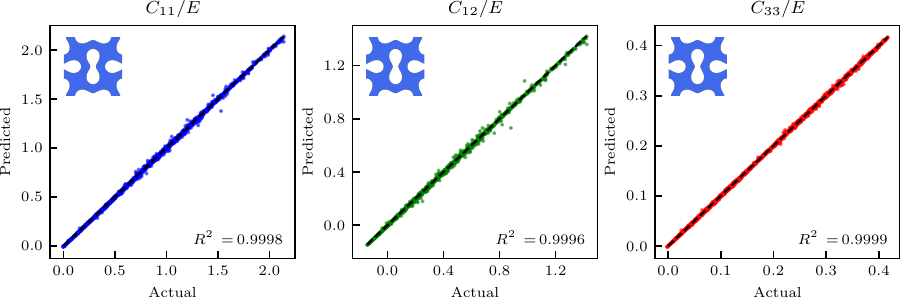}
            \caption{Test dataset}
            \label{fig:5b}
        \end{subfigure}
        \caption{Plots of the predicted values (obtained from the surrogate models) versus the actual values (obtained from the FFT-based solver) of the normalized effective elastic constants $C_{11}/E$, $C_{12}/E$, and $C_{33}/E$ for the peanut-shaped void unit cell. The diagonal dashed line depicts the perfect prediction.}
        \label{fig:5}
\end{figure}

The FFT-based homogenization scheme is used to generate datasets comprising $48k$ (i.e., $48000$) samples for each of the four unit cells. FFT-based simulations are performed on a $256\times256$ grid discretization of the geometry to compute the normalized effective elastic constants (through Eqs. \ref{eq:15} and \ref{eq:17}) using the MINRES solver with a tolerance of $10^{-6}$. The $48k$ data points are generated by randomly sampling from the input parameter space. This space is defined by three key parameters: Poisson's ratio of the base material $\nu$, ranging from $0.2$ to $0.4$, and the relative void parameters $d/L$ and $D/L$, both spanning from $0.0$ to $1.0$. An additional geometric constraint is imposed such that $d/L + D/L < 1$. This constraint ensures that the voids within the unit cell do not interconnect, maintaining the structural integrity of the unit cell. This comprehensive sampling strategy, incorporating both parameter ranges and geometric constraints, ensures that the surrogate models are trained on a diverse yet physically meaningful set of unit cell configurations.

It is worth highlighting that the utilization of such voluminous datasets is justified by the remarkable efficiency of the FFT-based solver, which enables the production of such large volumes of data in a relatively short period of time. To be more specific, it just took a maximum of around 8 hours of CPU time on a computing system equipped with two Intel Xeon Platinum 8160 processors, totaling $48$ physical cores (i.e., $2\times24$) to generate such a large dataset for each of the four unit cells. Moreover, employing a substantial dataset size is crucial for achieving highly accurate surrogate models that yield smooth solution fields (see Section \ref{sec:5.1}). Nonetheless, studies on the influence of dataset size on the accuracy of the surrogate models indicate that even significantly smaller datasets can produce models with reasonable accuracy. The effect of dataset size on the accuracy of the trained surrogate models is investigated in more detail in Section \ref{sec:5.1.1}.

For the training of the surrogate models, the following mean squared error $\mathcal{L}_{\mathrm{model}}$ is employed as the loss function:

\begin{equation}  
    \mathcal{L}_{\mathrm{model}} = \frac{1}{n_s} \sum_{i=1}^{n_s} (z_i - \hat{z}_i)^2,  
\end{equation}  

where $n_s$ denotes the number of samples, $z_i$ is the \(i\)-th actual value obtained from the FFT-based solver, and $\hat{z}_i$ is the $i$-th predicted value obtained from the trained model. In this context, the values $z_i$ and $\hat{z}_i$ specifically correspond to the normalized effective elastic constants $C_{11}/E$, $C_{12}/E$, or $C_{33}/E$, depending on the particular model being trained. Moreover, the performance of the models is evaluated using the coefficient of determination $R^2$, which is defined as  

\begin{equation}  
    R^2 = 1 - \frac{\sum_{i=1}^{n_s} (z_i - \hat{z}_i)^2}{\sum_{i=1}^{n_s} (z_i - \bar{z})^2}.  
\end{equation}  

Here, $\bar{z}$ is the mean of the actual values, which can be expressed by

\begin{equation}
    \bar{z} = \frac{1}{n_s} \sum_{i=1}^{n_s} z_i.
\end{equation}

To evaluate the predictive capability of the generated surrogate models, a train/test split methodology is employed. The dataset is randomly partitioned using a $90\%$/$10\%$ distribution, with $90\%$ of the data allocated for training and $10\%$ reserved for testing. Figures \ref{fig:2}--\ref{fig:5} illustrate the performance of the trained surrogate models in predicting the normalized effective elastic constants $C_{11}/E$, $C_{12}/E$, and $C_{33}/E$ for unit cells with rectangular, diamond, oval, and peanut-shaped voids, respectively. In each figure, subfigure (a) presents the results for the training set, while subfigure (b) shows the results for the test set. In these figures, the actual values and the corresponding predictions are plotted on the $x$ and $y$ axes, respectively, with the diagonal line (i.e., $x=y$) depicting the perfect prediction. Furthermore, to quantify the accuracy of the models, the coefficient of determination $R^2$ is employed as the primary criterion and prominently displayed up to four decimal places within the figures. As evident from these plots, the data points show minimal deviation from the diagonal, indicating remarkably high accuracy across all surrogate models, regardless of the unit cell geometry or the specific effective elastic constant being predicted. As anticipated, the models demonstrate superior performance on the training sets, given their prior exposure to this data. Nevertheless, the test sets also display promising accuracy despite being unseen, with $R^2$ values no lower than 0.9993 for all cases. This consistently high level of performance across both seen and unseen data, and across all unit cell types and elastic constants, underscores the robustness and generalizability of the generated surrogate models.

Among different unit cell types, the diamond void unit cell exhibits the best performance, as evidenced by both the minimal deviation of data points from the diagonal line and the highest coefficient of determination $R^2$. In contrast, while still maintaining solid predictive capabilities, the oval void unit cell shows slightly more variability in its predictions. Nevertheless, it is noteworthy that even for the oval void models, the $R^2$ values do not fall under 0.9993, highlighting the consistently high precision of the generated surrogate models across all four unit cell geometries. When comparing the performance of the surrogate models for different effective elastic constants, the predictions for $C_{33}/E$ consistently achieve the highest accuracy across all unit cell types. Conversely, while $C_{12}/E$ predictions show the largest deviations from the ideal line, the generated models still achieve reliable accuracy with $R^2$ values no lower than 0.9993. This demonstrates the overall reliability of the generated surrogate models for all three effective elastic constants.

\section{Inverse analysis}
\label{sec:4}

The possibility of evaluating the generated surrogate models in real time enables exploring the design space and finding the specific geometric and material input parameters that result in the desired effective stiffness components. For instance, see Section \ref{sec:6}, where even an inexperienced user can probe the design space in real time using a newly developed GUI.

While the real-time evaluation of the surrogate models is highly advantageous, establishing an inverse analysis framework utilizing the generated surrogate models remains highly beneficial. In such inverse approaches, the input parameters that yield the desired outputs are determined through iterative schemes, such as gradient-based or heuristic optimization methods. However, the real-time evaluation of the generated surrogate models provides the opportunity of using a brute-force algorithm to swiftly solve the inverse problem.

In a brute-force optimization algorithm, the objective function is evaluated at every possible combination of input parameters within the specified search space to identify the optimal set that meets the desired criteria. This approach ensures finding the global optimum (if it exists) while avoiding the problem of entrapment in local minima, which is a common challenge faced by gradient-based algorithms. It is also worth noting that increasing the number of points at which the surrogate models are evaluated improves the precision of the obtained solutions. This is investigated in more detail through a numerical example in Section \ref{sec:5.2}.  

In the current work, considering that the base material for the fabrication of the structures, and thereby its Poisson's ratio $\nu$, is known, the inverse problems involves finding the optimal relative void diameters $d/L$ and $D/L$ that result in the desired values for the components of the normalized effective tangent stiffness $\hat{y}_i$, defined as

\begin{equation}
\hat{y}_1=\frac{\hat{C}_{11}}{E}, \quad \hat{y}_2=\frac{\hat{C}_{12}}{E}, \quad \hat{y}_3=\frac{\hat{C}_{33}}{E}.
\end{equation}

Here, $\hat{(\cdot)}$ is used to denote the desired output values. The predicted values $y_i$ are then obtained by evaluating the corresponding surrogate models $S_i$ for each component of the effective stiffness matrix as 

\begin{equation}
y_1=\frac{C_{11}}{E}=S_1\left(\frac{d}{L}, \frac{D}{L}, \nu\right), \quad y_2=\frac{C_{12}}{E}=S_2\left(\frac{d}{L}, \frac{D}{L}, \nu\right), \quad y_3=\frac{C_{33}}{E}=S_3\left(\frac{d}{L}, \frac{D}{L}, \nu\right).
\end{equation}

Then, by computing the mean squared error $\mathcal{L}_{\mathrm{inv}}$ between the predicted values $y_i$ and the desired output values $\hat{y}_i$ using

\begin{equation}
    \mathcal{L}_{\mathrm{inv}} \left(\frac{d}{L}, \frac{D}{L}, \nu \right)= \frac{1}{n_c} \sum_{i=1}^3 w_i\left(y_i-\hat{y}_i\right)^2,
\end{equation}

the optimal set of input parameters $d/L$ and $D/L$ resulting in an error smaller than a chosen threshold can be extracted. In the above, $n_c=\sum_{i=1}^3 w_i$, where $w_i$ is a boolean variable indicating whether a particular component or any combination of the components of the effective stiffness matrix (7 possible combinations) is specified as the desired output, and it is defined as follows:

\begin{equation}
w_i= \begin{cases}1 & \text { if } \hat{y}_i \text { is specified } \\ 0 & \text { if } \hat{y}_i \text { is not specified }\end{cases}
\end{equation}

When specifying the desired output as a combination of the components of the effective stiffness matrix, it is important to note that the solution to the inverse problem for such a combination may or may not exist within the defined error threshold. Conversely, if only a particular component is specified as the desired output, the likelihood of the inverse problem having a solution is higher, provided that the desired value is both physical and possible. A practical example with specifying two components of the effective tangent stiffness as the desired output can be followed in Section \ref{sec:5.2} for the rectangular void unit cell.

\section{Case studies}
\label{sec:5}

This section presents two comprehensive case studies to demonstrate the capabilities and applications of the developed surrogate models for auxetic metamaterials. Firstly, Section \ref{sec:5.1} details a parametric study, which serves to examine the accuracy of the surrogate models and explore the auxetic behavior of the unit cells across a range of geometric configurations. Following this, Section \ref{sec:5.2} addresses an inverse problem, highlighting the practical applicability of the developed surrogate models in tailoring auxetic metamaterials with desired mechanical properties.

\subsection{A parametric study}
\label{sec:5.1}

In this section, a parametric study is performed to both evaluate the accuracy of the developed surrogate models and examine the auxetic behavior of the unit cells for different geometric parameters. To this end, by fixing the Poisson's ratio of the base material as $\nu=0.3$ and one of the relative void diameters as $d/L=0.05$, the other relative void diameter $D/L$ is changed within the interval $[0.05,0.9]$ with a step size of $0.01$.

\begin{figure}[h!]
    \centering
    \includegraphics[width=\linewidth, keepaspectratio]{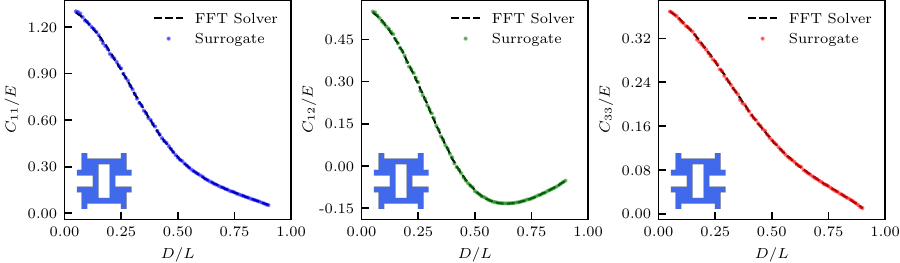}
    \caption{Plots of the normalized effective elastic constants $C_{11}/E$, $C_{12}/E$, and $C_{33}/E$ versus the relative void diameter $D/L$ for the rectangular void unit cell. Poisson's ratio of the base material and the other relative void diameter are fixed as $\nu=0.3$ and $d/L=0.05$, respectively.}
    \label{fig:6}
\end{figure}

\begin{figure}[h!]
    \centering
    \includegraphics[width=\linewidth, keepaspectratio]{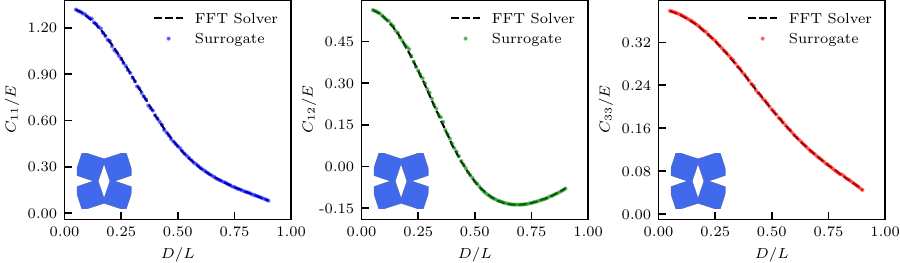}
    \caption{Plots of the normalized effective elastic constants $C_{11}/E$, $C_{12}/E$, and $C_{33}/E$ versus the relative void diameter $D/L$ for the diamond void unit cell. Poisson's ratio of the base material and the other relative void diameter are fixed as $\nu=0.3$ and $d/L=0.05$, respectively.}
    \label{fig:7}
\end{figure}

\begin{figure}[h!]
    \centering
    \includegraphics[width=\linewidth, keepaspectratio]{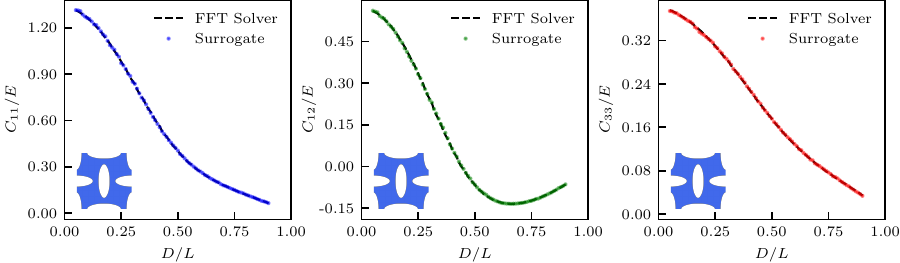}
    \caption{Plots of the normalized effective elastic constants $C_{11}/E$, $C_{12}/E$, and $C_{33}/E$ versus the relative void diameter $D/L$ for the oval void unit cell. Poisson's ratio of the base material and the other relative void diameter are fixed as $\nu=0.3$ and $d/L=0.05$, respectively.}
    \label{fig:8}
\end{figure}

\begin{figure}[h!]
    \centering
    \includegraphics[width=\linewidth, keepaspectratio]{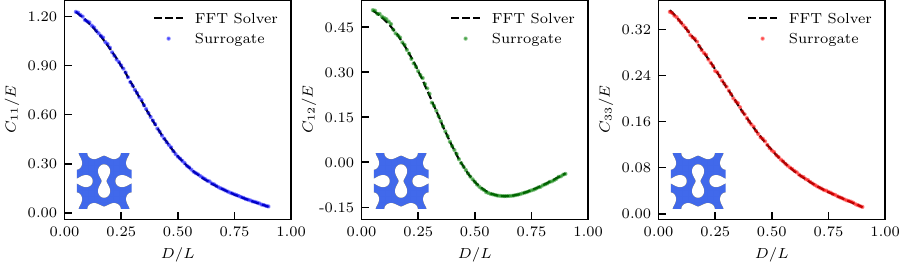}
    \caption{Plots of the normalized effective elastic constants $C_{11}/E$, $C_{12}/E$, and $C_{33}/E$ versus the relative void diameter $D/L$ for the peanut-shaped void unit cell. Poisson's ratio of the base material and the other relative void diameter are fixed as $\nu=0.3$ and $d/L=0.05$, respectively.}
    \label{fig:9}
\end{figure}

The normalized effective elastic constants $C_{11}/E$, $C_{12}/E$, and $C_{33}/E$ are plotted versus the relative void diameter $D/L$ in Figures \ref{fig:6}--\ref{fig:9} for all four unit cells. It is observed that for all three effective elastic constants and all four structures, the surrogate models provide results in close agreement with those obtained from the FFT-based solver. The curves exhibit a consistent overall trend, indicating a general similarity in the effect of the void size (i.e., $D/L$) on the effective elastic constants, whereas subtle differences are apparent in the slopes of these curves. Interestingly, despite the slight differences in the slopes, the maximum and minimum values of each effective elastic constant remain comparable across different auxetic unit cells.

\begin{figure}[h!]
    \centering
    \includegraphics[width=0.75\linewidth, keepaspectratio]{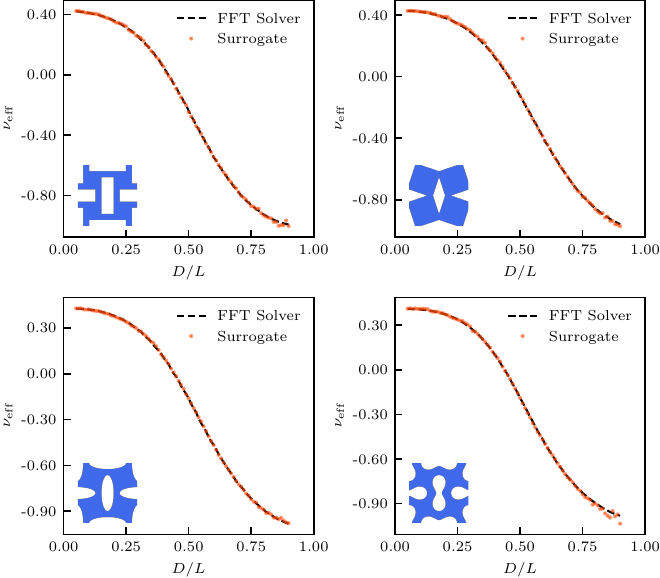}
    \caption{Plots of the effective Poisson's ratio $\nu_{\mathrm{eff}}$ versus the relative void diameter $D/L$ for the auxetic unit cells with rectangular, diamond, oval, and peanut-shaped voids. The Poisson's ratio of the base material and the other relative void diameter are fixed as $\nu=0.3$ and $d/L=0.05$, respectively.}
    \label{fig:10}
\end{figure}

The auxeticity of the structures is identified from the negative values of $C_{12}$ in Figures \ref{fig:6}--\ref{fig:9} for a specific range of the void diameter $D/L$. In order to provide a clearer perspective on the auxeticity of the structures for different void sizes, the effective Poisson's ratio is computed as $\nu_\mathrm{eff} = C_{12}/C_{11}$ and plotted versus the relative void diameter $D/L$ in Figure \ref{fig:10} for all four unit cells. It is observed that for the unit cells with rectangular, diamond, oval, and peanut-shaped voids, the structures enter the auxetic region for the relative void diameters $D/L$ equal and greater than $0.42$, $0.46$, $0.45$, and $0.44$, respectively. It is worth highlighting that both the FFT-based solver and surrogate models yield identical values of the aforementioned critical void diameters, thereby demonstrating the high accuracy of the generated surrogate models. However, it can be seen that for the rectangular void and peanut-shaped void unit cells, the accuracy of the surrogate models slightly decreases when the relative void diameter $D/L$ approaches $0.9$. This can be well justified by the fact that by increasing the void diameter, the effective elastic constants approach zero as the amount of material decreases in the structure (see Figures \ref{fig:6}--\ref{fig:9}). For such large values of the relative void diameter $D/L$, the effective Poisson's ratio $\nu_{\mathrm{eff}}=C_{12}/C_{11}$ is obtained by dividing two small numbers close to zero. Therefore, even a small error in each of the elastic constants $C_{12}$ and $C_{11}$ may lead to a higher error in their fraction. In case there is a need for highly accurate solutions for the effective Poisson's ratio $\nu_{\mathrm{eff}}$, a separate surrogate model to predict this value can be generated as well.

\subsubsection{Dataset size}
\label{sec:5.1.1}

As mentioned in Section \ref{sec:3.3}, the surrogate models are generated from datasets of size $48k$. The reasons to consider such large datasets are twofold: Firstly, the utilized FFT-based solver is so efficient that generating such a large volume of data is possible within a few hours (i.e., around 8 hours of CPU time with 48 core processors; see Section \ref{sec:3.3}). Secondly, utilizing a large volume of data ensures obtaining highly accurate surrogate models that can provide smooth solution fields, such as those presented in Figures \ref{fig:6}--\ref{fig:9}.

Nevertheless, investigating the effect of dataset size on the accuracy of the surrogate models reveals that even significantly smaller datasets can still generate adequately accurate surrogate models. To explain more thoroughly, surrogate models for the $C_{11}$ component of the effective elastic stiffness are generated using dataset sizes of $3k$, $6k$, $12k$, and $24k$ in addition to the already available models with $48k$ data points. Then, these new surrogate models are utilized for the same problem of Section \ref{sec:5.1} to compute the normalized elastic constant $C_{11}/E$ for all four unit cells. Completely similar to the previous parametric study (Section \ref{sec:5.1}), the models are evaluated for fixed values of $\nu=0.3$ and $d/L=0.05$ while $D/L$ is varied in the range $[0.05,0.9]$ with a step size of $0.01$.

\begin{figure}[ht]
    \centering
    \includegraphics[width=0.75\linewidth, keepaspectratio]{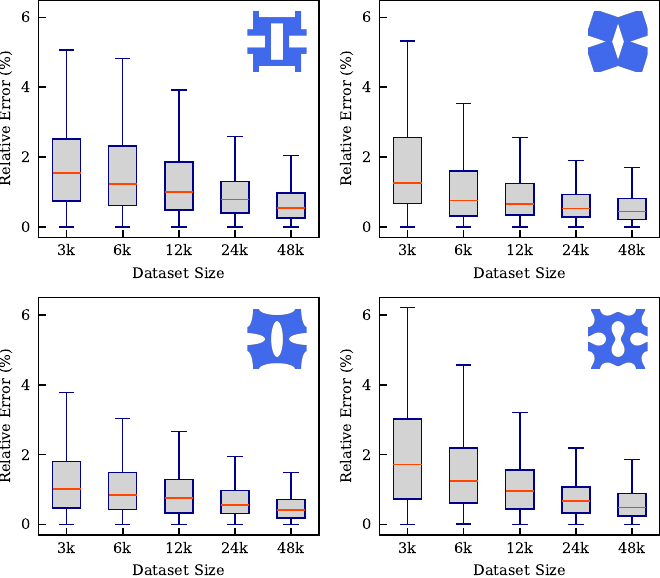}
    \caption{Box plots for the distribution of the relative error in the normalized effective elastic constant $C_{11}/E$ between the actual values obtained from the FFT-based solver and the predicted values obtained from the surrogates models trained on varying dataset sizes. Results are presented for the auxetic unit cells with rectangular, diamond, oval, and peanut-shaped voids.}
    \label{fig:11}
\end{figure}

The relative errors between the values obtained from the surrogate models trained on varying dataset sizes and the FFT-based solver are computed for all the values of relative void diameter $D/L$ and presented as box plots in Figure \ref{fig:11} for all four unit cells. A noticeable trend is the reduction in the median relative error as the dataset size increases, indicating an overall improvement in model accuracy. Additionally, the interquartile range (IQR) becomes narrower, highlighting a decrease in the variability of the relative error. The length of the whiskers also reduces, suggesting a tightening in the overall spread of the error distribution. It is worth noting that, although utilizing a large dataset consisting of $48k$ data points can result in higher precision with a relative error distribution below $2 \, \%$, even significantly reduced datasets, such as a dataset consisting of $3k$ data points, are capable of producing reasonably accurate surrogate models with a relative error distribution around or below $6\, \%$.

Regarding the performance of surrogate models trained on the smallest dataset size (i.e., $3k$ data points) for different unit cell types, a comparison of the box plots reveals that the oval void unit cell achieves the highest accuracy with smaller datasets, exhibiting a relative error distribution below $4 \, \%$. In contrast, the peanut-shaped void unit cell shows the highest error, slightly exceeding $6 \, \%$. The rectangular and diamond void unit cells demonstrate similar performance, with errors just below $6 \, \%$. However, for all four unit cells, using the dataset with $48k$ data points results in highly accurate models with relative errors below $2 \, \%$.

\subsection{An inverse problem}
\label{sec:5.2}

In this section, an inverse problem is solved to examine the performance of the generated surrogate models for inverse design. It is considered that the base material for fabricating the structure is the additively manufactured poly-lactic acid (PLA) with the elastic properties $E=3500 \, \mathrm{MPa}$ and $\nu=0.36$ chosen from \cite{PLAprop}. The goal here is to obtain the relative void diameters $d/L$ and $D/L$ for the rectangular void unit cell that yield the desired effective stiffness components $C_{11}=400 \, \mathrm{MPa}$ and $C_{12}=-200 \, \mathrm{MPa}$. Such values of the effective stiffness components result in a negative effective Poisson's ratio $\nu_{\mathrm{eff}} = C_{12}/C_{11} = -0.5$.

The inverse problem is solved using the brute-force algorithm described in Section \ref{sec:4}. As explained in Section \ref{sec:4}, the accuracy of the solutions obtained for the inverse problem depends on the number of evaluation points. To investigate this relationship, Table \ref{tab:1} presents the relative void diameters $d/L$ and $D/L$ obtained from the inverse problems using $1k$ to $20k$ evaluation points, along with their corresponding CPU times. As evident, the solution of the inverse problem converges as the number of evaluation points increases. The relative void diameters $d/L$ and $D/L$ show a trend towards convergence, reaching values of 0.34 and 0.50, respectively. Notably, even with such a large number of evaluation points as $20k$, the inverse problem is still solved efficiently, with the CPU time remaining below 1 second.

\begin{table}[h!]
\caption{Relative void diameters $d/L$ and $D/L$ obtained from inverse problems using the brute-force algorithm with varying numbers of evaluation points from $1k$ to $20k$ and the corresponding CPU times.}
\label{tab:1}
\centering
    \begin{tabular}{@{}cccc@{}}
    \toprule
    Evaluation count & $d/L$ & $D/L$ & CPU time (s) \\ \midrule
    $1k$       & $0.32$     & $0.51$       &   $0.11$    \\
    $2.5k$     & $0.33$     & $0.51$         &   $0.21$   \\
    $5k$       & $0.33$     & $0.50$       &   $0.33$   \\ 
    $10k$      & $0.34$     & $0.50$       &   $0.56$   \\
    $20k$      & $0.34$     & $0.50$       &   $0.99$   \\ \bottomrule
\end{tabular}
\end{table}

\begin{figure}[h!]  
    \centering
    \begin{subfigure}{0.4\textwidth}  
        \centering  
        \includegraphics[width=\textwidth]{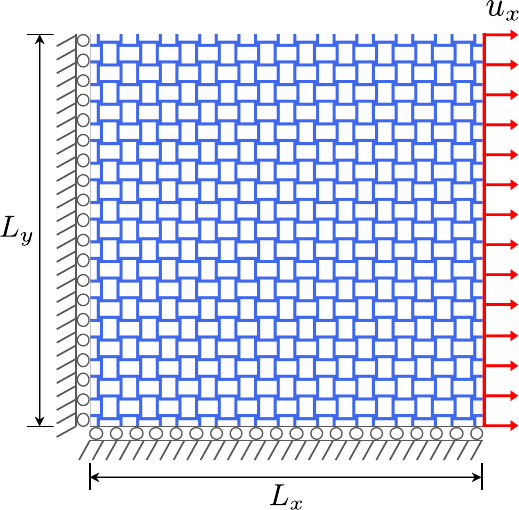}  
        \caption{First case: computation of $\nu_{\mathrm{eff}}$}  
        \label{fig:12a}  
    \end{subfigure}  
    \hspace{0.02\textwidth} 
    \begin{subfigure}{0.05\textwidth}  
        \centering  
        \includegraphics[width=\textwidth]{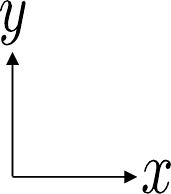}  
        \label{fig:subfig1}  
    \end{subfigure}  
    \hspace{0.02\textwidth} 
    \begin{subfigure}{0.4\textwidth}  
        \centering  
        \includegraphics[width=\textwidth]{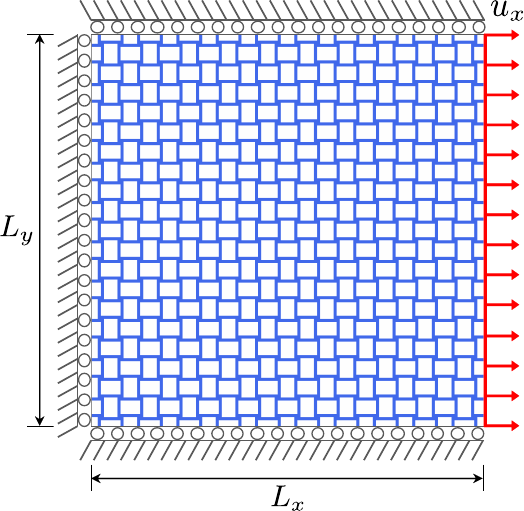}  
        \caption{Second case: computation of $C_{11}$ and $C_{12}$}  
        \label{fig:12b}  
    \end{subfigure}  
    \caption{Geometry and boundary conditions for the specimens made form a $10\times10$ distribution of the rectangular void unit cell with geometric parameters obtained form the inverse problem. In both cases, the left and bottom boundaries are fixed in the $x$ and $y$ directions, respectively, and a displacement of $u_x=1 \, \mathrm{mm}$ is applied to the right boundary. This is while, the top boundary is free for the first case (subfigure \textbf{a}) and fixed in the $y$ direction for the second case (subfigure \textbf{b}).}  
    \label{fig:12}  
\end{figure}

After obtaining the values $d/L=0.34$ and $D/L=0.50$ from the inverse problem, a $10 \times 10$ distribution of the rectangular void unit cell with a side length of $10 \, \mathrm{mm}$ is simulated with FEM to examine the accuracy of the obtained design parameters. As shown in Figure \ref{fig:12}, two different sets of boundary conditions are considered to obtain $\nu$, $C_{11}$ and $C_{12}$ from FE simulations.

In the first case (Figure \ref{fig:12a}), the left and bottom boundaries of the specimen are fixed in the $x$ and $y$ directions, respectively, and a displacement of $u_x=1 \, \mathrm{mm}$ is applied to the right boundary. Then, by computing the displacement of the transverse boundary, i.e., the top boundary, the effective Poisson's ratio can be computed by

\begin{equation}
\label{eq:73}
    \nu_{\mathrm{eff}} = - \dfrac{\bar{\varepsilon}_{yy}}{\bar{\varepsilon}_{xx}} \qquad \mathrm{with} \quad \bar{\varepsilon}_{xx} = \dfrac{u_{x}}{L_{x}}, \quad \bar{\varepsilon}_{yy} = \dfrac{u_{y}}{L_{y}},
\end{equation}

where the effective strains of the entire specimen in the $x$ and $y$ directions are represented by $\bar{\varepsilon}_{xx}$ and $\bar{\varepsilon}_{yy}$, respectively. These strains are related to the displacements of the specimen's boundaries: $u_x$ refers to the displacement of the right boundary in the $x$ direction, while $u_y$ corresponds to the displacement of the top boundary in the $y$ direction. Moreover, $L_x$ and $L_y$ denote the widths of the specimen in the $x$ and $y$ directions, respectively.

In the second case (Figure \ref{fig:12b}), the boundary conditions are similar to those in the first case but the top boundary is additionally fixed in the $y$ direction. Here, the effective stiffness components $C_{11}$ and $C_{12}$ can be computed by

\begin{equation}
    C_{11} = \frac{\bar{\sigma}_{xx}}{\bar{\varepsilon}_{xx}}, \quad  C_{12} = \frac{\bar{\sigma}_{yy}}{\bar{\varepsilon}_{xx}} \qquad \mathrm{with} \quad \bar{\sigma}_{xx} = \frac{f_x}{tL_y}, \quad \bar{\sigma}_{yy} = \frac{f_y}{tL_x},
\end{equation}

where $t$ is the thickness of the specimen. Furthermore, the sum of reaction forces in the $x$ direction on the right boundary is denoted by $f_x$, while $f_y$ represents the sum of reaction forces in the $y$ direction on the top boundary.

The obtained values from FE simulations and the corresponding error with respect to  the desired values are summarized in Table \ref{tab:2}. It can be seen that the effective Poisson's ratio $\nu_{\mathrm{eff}}$ shows an exact match between the desired and obtained values up to three decimal places, resulting in a zero error. For the effective elastic constants $C_{11}$ and $C_{12}$, the obtained values from FEM deviate slightly from the desired values with percentage errors of $0.67 \, \%$ and $0.06 \, \%$, respectively, indicating a high degree of accuracy in the obtained design parameters from the inverse problem.

\begin{table}[h!]
\caption{Comparison of the obtained values from FE simulations using the geometric parameters achieved from the inverse problem for the rectangular void unit cell against the desired values and their corresponding relative error.}
\label{tab:2}
\centering
    \begin{tabular}{@{}cccc@{}}
    \toprule
    Quantity & Desired value & Obtained value (FEM) & Error \\ \midrule
    $\nu_{\mathrm{eff}}\, [-]$    & $-0.500$     & $-0.500$       &   $0.00 \, \%$    \\
    $C_{11}\, [\mathrm{MPa}]$                & $400.00$     & $402.69$       &   $0.67 \, \%$   \\
    $C_{12}\, [\mathrm{MPa}]$                & $-200.00$    & $-199.89$      &   $0.06 \, \%$   \\ \bottomrule
\end{tabular}
\end{table}

\section{Graphical user interface (GUI)}
\label{sec:6}

\begin{figure}[h!]  
    \centering  
    \begin{subfigure}{1\textwidth}  
        \centering  
        \includegraphics[width=0.75\textwidth, keepaspectratio]{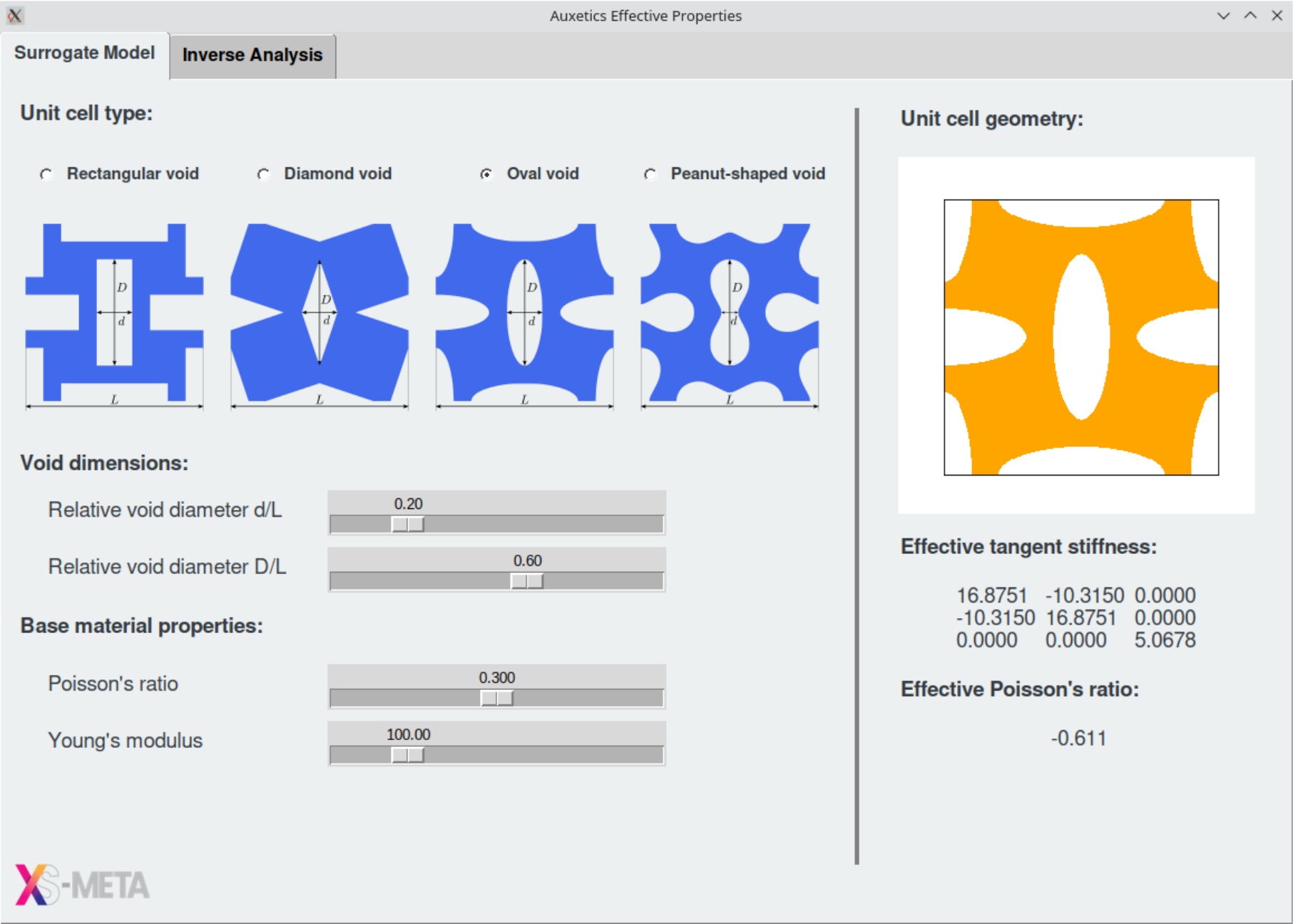}  
        \caption{Surrogate Model tab for the real-time prediction of the effective tangent stiffness.}  
        \label{fig:13a}  
    \end{subfigure}  
    \vspace{0.05\textwidth}
    
    \begin{subfigure}{1\textwidth}  
        \centering  
        \includegraphics[width=0.75\textwidth, keepaspectratio]{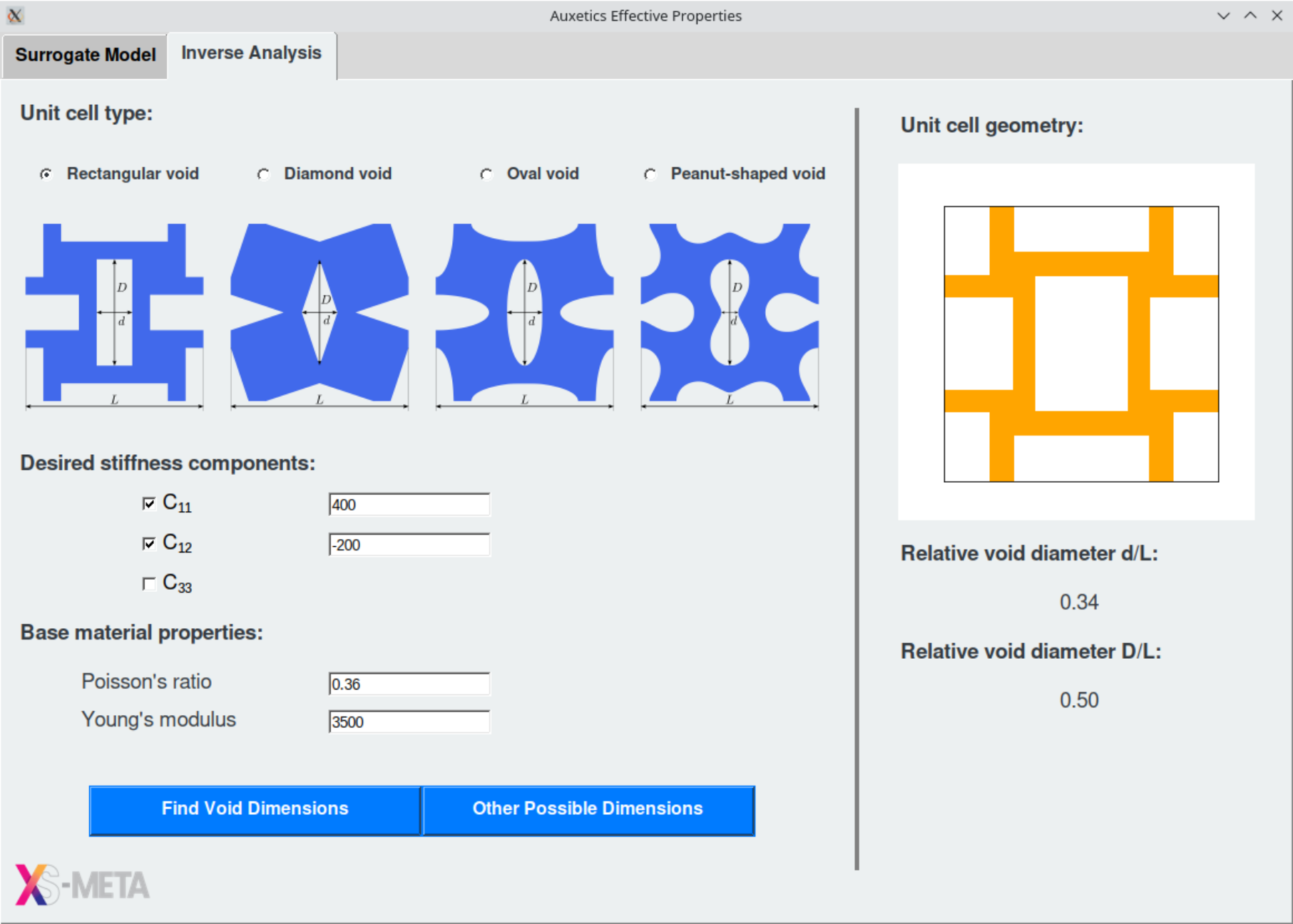}  
        \caption{Inverse Analysis tab for finding the relative void diameters $d/L$ and $D/L$.}  
        \label{fig:13b}  
    \end{subfigure}  
    \caption{GUI for the real-time prediction of the effective tangent stiffness and the inverse analysis of the auxetic unit cells.}  
    \label{fig:13}  
\end{figure}

To facilitate the practical usage of the generated surrogate models for less experienced users, a GUI has been developed using Python's Tkinter library. The developed GUI consists of two tabs, including the \textbf{Surrogate Model} tab (Figure \ref{fig:13a}) and the \textbf{Inverse Analysis} tab (Figure \ref{fig:13b}).

In the Surrogate Model tab, users can extract the effective tangent stiffness $\mathbb{C}$ and the resulting effective Poisson's ratio $\nu_{\mathrm{eff}}$ in real time for the desired auxetic structure. Users can change the geometric input parameters $d/L$ and $D/L$ as well as the material input parameters $E$ and $\nu$. Moreover, the unit cell geometry is plotted in real time for users. 

In the Inverse Analysis tab, users can extract the relative void diameters $d/L$ and $D/L$ for a selected unit cell, given the desired components of the effective elastic stiffness and the base material properties $E$ and $\nu$. Additionally, the corresponding unit cell geometry with the obtained geometric parameters is plotted. It is worth mentioning that, as detailed in Section \ref{sec:4}, users are given the flexibility to select any component or any combination of the components of the effective elastic stiffness as the target output. However, it is important to recognize the inherent constraints of the inverse problem. Users should be reminded that not all values of desired effective elastic properties are achievable within the physical limitations of the unit cell design. The space of all possible solutions is bounded by the base material properties and geometric constraints of the unit cell structure. Therefore, if the desired values lie outside this feasible region, the inverse analysis fails to provide a valid solution. Users are advised to consider these physical limitations when setting the desired outputs to ensure that the inverse problem remains solvable.

\section{Conclusions}
\label{sec:7}

The present study focused on developing surrogate models for real-time prediction of the effective elastic properties of auxetic unit cells with orthogonal voids of different shapes, including rectangular, diamond, oval, and peanut-shaped voids. These surrogate models accept the relative void diameters $d/L$ and $D/L$ along with Poisson's ratio of the base constituent material $\nu$ as input parameters and deliver the normalized effective elastic constants $C_{11}/E$,  $C_{12}/E$, and $C_{33}/E$ as their outputs. These three constants represent the only distinct values necessary to fully define the homogenized effective elastic stiffness tensor.

The FFT-based homogenization scheme was employed to efficiently generate the datasets required for training the surrogate models. This particular solver facilitated the rapid generation of extensive datasets within a relatively short time frame, circumventing the challenges typically associated with FEM. More specifically, it eliminated the need for careful consideration of symmetrical meshes at the unit cell boundaries and the application of periodic boundary conditions, which are often cumbersome in FE approaches. The efficiency of this approach was demonstrated by the generation of a comprehensive dataset comprising $48k$ data points, accomplished within a maximum of 8 hours of CPU time using a computer with $48$ CPU cores. 

The random forests regression method was used to develop surrogate models based on the datasets generated by the FFT-based solver. The performance of the generated surrogate models was examined through a train/test split methodology. By comparing the predicted values obtained from the models with the actual values obtained from the FFT-based solver, it was demonstrated that the generated surrogate models are capable of achieving a high level of accuracy in the prediction of the effective elastic constants even for the test set, which contained unseen data. In addition, a parametric study was also conducted to evaluate the accuracy of the developed surrogate models and investigate the auxetic behavior of the unit cells across diverse geometric configurations. This analysis not only showcased the high precision of the surrogate models but also provided valuable insights into the relationship between structural geometry and auxetic properties. The study investigated the impact of dataset size on the accuracy of the surrogate models, revealing that even substantially reduced datasets can produce satisfactory results.

An inverse design framework has been established to determine the geometric parameters that yield desired effective stiffness constants. This inverse problem was solved using a brute-force algorithm, which evaluates the surrogate models at a predefined number of points in the parameter space to extract the optimal set satisfying the desired requirements. The rapid evaluation of the surrogate models makes this approach feasible without the risk of becoming trapped in local minima, resulting in a practical and robust solution method. To assess the capability of the established surrogate models in inverse design, a practical inverse problem was solved for the rectangular void unit cell. Investigations revealed that the solution to the inverse problem converges as the number of evaluation points for the brute-force algorithm increases. Even using a high number of evaluation points (i.e., $20k$ points) does not cause the inverse analysis time to exceed $1$ second. FE simulations were performed using the obtained geometric parameters from the inverse problem to examine the performance. It was shown that the obtained values from FE simulations, i.e., the effective Poisson's ratio and the effective elastic constants, were in close agreement with the target values for the inverse problem with a maximum error of $0.67\,\%$. This underscores the benefit of the generated surrogate models in designing auxetic structures with tailored mechanical properties. 

A user-friendly GUI has also been developed to enhance the accessibility and practical application of surrogate models for auxetic metamaterials. The GUI features two main components: a tab for real-time prediction of effective elastic constants based on user-defined geometric and material inputs, and a second tab for determining the optimal unit cell geometry to achieve desired effective elastic properties. This tool not only facilitates rapid effective properties prediction and inverse design but also provides visual feedback through real-time unit cell geometry plotting.

Moving forward, this study can be extended in several directions for subsequent research. Developing surrogate models capable of predicting full-field strain/stress distributions would be valuable for applications where such detailed information is crucial. These models would incorporate macroscopic strain field values in addition to geometric and material input parameters. Moreover, creating surrogate models for the finite deformation behavior of both auxetic and non-auxetic lattice structures could significantly reduce computational effort in two-scale homogenization approaches. Such models would take the macroscopic deformation gradient as input and yield the corresponding homogenized tangent stiffness at each macroscopic integration point during incremental loading. Eventually, another avenue for future research is to establish an active learning framework with on-the-fly data generation, in which the FFT-based solver is used to generate data points dynamically during the training process. This approach would enhance the accuracy of the generated surrogate models and the quality of fit, while avoiding the need to generate an extensive dataset prior to the training stage, potentially leading to more efficient and adaptive model development.

\section*{Acknowledgments}

\begin{minipage}[b]{2.cm}
	\begin{figure}[H] %
		\centering
    \includegraphics[width=0.6\linewidth]{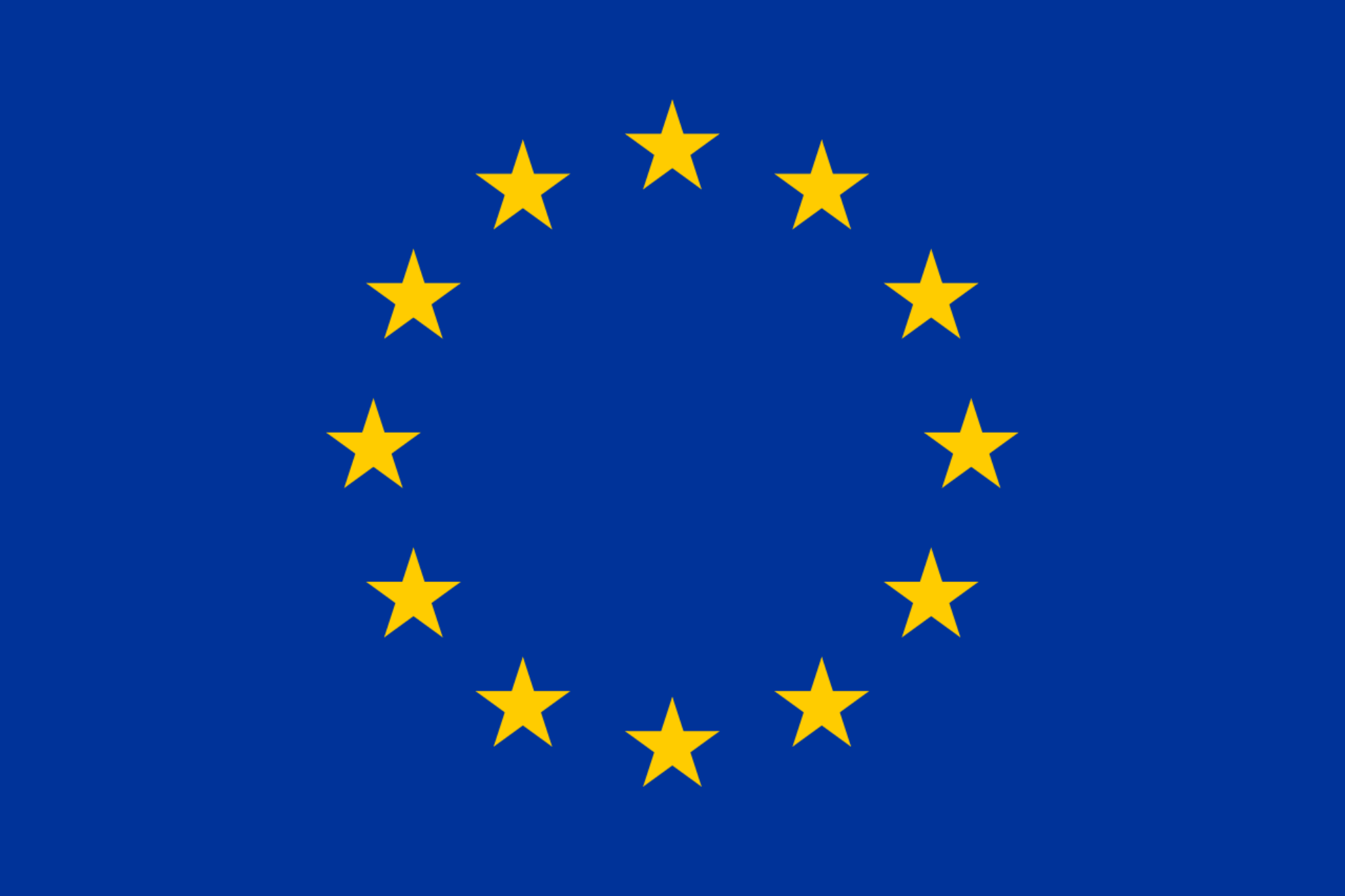}
	\end{figure}    
\end{minipage}
\begin{minipage}{15.5cm}
    The authors gratefully acknowledge the funding from the European Union’s Horizon 2020 research and innovation program under the Marie Skłodowska-Curie grant agreement No 956401 (XS-Meta). 
\end{minipage}


\printbibliography

@article{zhang2020large,
  title={Large deformation and energy absorption of additively manufactured auxetic materials and structures: A review},
  author={Zhang, Jianjun and Lu, Guoxing and You, Zhong},
  journal={Composites Part B: Engineering},
  volume={201},
  pages={108340},
  year={2020},
  publisher={Elsevier}
}

@article{najafi2021experimental,
  title={Experimental investigation on energy absorption of auxetic structures},
  author={Najafi, Milad and Ahmadi, Hamed and Liaghat, Gholamhossein},
  journal={Materials today: proceedings},
  volume={34},
  pages={350--355},
  year={2021},
  publisher={Elsevier}
}

@article{choudhry2022plane,
  title={In-plane energy absorption characteristics of a modified re-entrant auxetic structure fabricated via 3D printing},
  author={Choudhry, Niranjan Kumar and Panda, Biranchi and Kumar, S},
  journal={Composites Part B: Engineering},
  volume={228},
  pages={109437},
  year={2022},
  publisher={Elsevier}
}

@article{yan2019novel,
  title={A novel cellular substrate for flexible electronics with negative Poisson ratios under large stretching},
  author={Yan, ZG and Wang, BL and Wang, KF and Zhang, Chunwei},
  journal={International Journal of Mechanical Sciences},
  volume={151},
  pages={314--321},
  year={2019},
  publisher={Elsevier}
}

@article{jiang2022flexible,
  title={Flexible metamaterial electronics},
  author={Jiang, Shan and Liu, Xuejun and Liu, Jianpeng and Ye, Dong and Duan, Yongqing and Li, Kan and Yin, Zhouping and Huang, YongAn},
  journal={Advanced Materials},
  volume={34},
  number={52},
  pages={2200070},
  year={2022},
  publisher={Wiley Online Library}
}

@article{jang2022auxetic,
  title={Auxetic meta-display: stretchable display without image distortion},
  author={Jang, Bongkyun and Won, Sejeong and Kim, Jaegu and Kim, Juho and Oh, Minsub and Lee, Hak-Joo and Kim, Jae-Hyun},
  journal={Advanced functional materials},
  volume={32},
  number={22},
  pages={2113299},
  year={2022},
  publisher={Wiley Online Library}
}

@article{allen2015auxetic,
  title={Auxetic foams for sport safety applications},
  author={Allen, Tom and Martinello, Nicolo and Zampieri, Davide and Hewage, Trishan and Senior, Terry and Foster, Leon and Alderson, Andrew},
  journal={Procedia Engineering},
  volume={112},
  pages={104--109},
  year={2015},
  publisher={Elsevier}
}

@article{duncan2018review,
  title={Review of auxetic materials for sports applications: Expanding options in comfort and protection},
  author={Duncan, Olly and Shepherd, Todd and Moroney, Charlotte and Foster, Leon and Venkatraman, Praburaj D and Winwood, Keith and Allen, Tom and Alderson, Andrew},
  journal={Applied Sciences},
  volume={8},
  number={6},
  pages={941},
  year={2018},
  publisher={MDPI}
}

@article{tahir2022auxetic,
  title={Auxetic materials for personal protection: A review},
  author={Tahir, Danish and Zhang, Minglonghai and Hu, Hong},
  journal={physica status solidi (b)},
  volume={259},
  number={12},
  pages={2200324},
  year={2022},
  publisher={Wiley Online Library}
}

@article{kolken2018rationally,
  title={Rationally designed meta-implants: a combination of auxetic and conventional meta-biomaterials},
  author={Kolken, Helena MA and Janbaz, Shahram and Leeflang, Sander MA and Lietaert, Karel and Weinans, Harrie H and Zadpoor, Amir A},
  journal={Materials Horizons},
  volume={5},
  number={1},
  pages={28--35},
  year={2018},
  publisher={Royal Society of Chemistry}
}

@article{kolken2020mechanical,
  title={Mechanical performance of auxetic meta-biomaterials},
  author={Kolken, HMA and Lietaert, K and Van der Sloten, T and Pouran, B and Meynen, A and Van Loock, G and Weinans, Harrie and Scheys, L and Zadpoor, AA},
  journal={journal of the mechanical behavior of biomedical materials},
  volume={104},
  pages={103658},
  year={2020},
  publisher={Elsevier}
}

@article{shirzad2023auxetic,
  title={Auxetic metamaterials for bone-implanted medical devices: recent advances and new perspectives},
  author={Shirzad, Masoud and Zolfagharian, Ali and Bodaghi, Mahdi and Nam, Seung Yun},
  journal={European Journal of Mechanics-A/Solids},
  volume={98},
  pages={104905},
  year={2023},
  publisher={Elsevier}
}

@article{masters1996models,
  title={Models for the elastic deformation of honeycombs},
  author={Masters, IG and Evans, KE},
  journal={Composite structures},
  volume={35},
  number={4},
  pages={403--422},
  year={1996},
  publisher={Elsevier}
}

@article{chan1999mechanical1,
  title={The mechanical properties of conventional and auxetic foams. Part I: compression and tension},
  author={Chan, N and Evans, KE},
  journal={Journal of Cellular plastics},
  volume={35},
  number={2},
  pages={130--165},
  year={1999},
  publisher={Sage Publications Sage CA: Thousand Oaks, CA}
}

@article{chan1999mechanical2,
  title={The mechanical properties of conventional and auxetic foams. Part II: shear},
  author={Chan, N and Evans, KE},
  journal={Journal of cellular plastics},
  volume={35},
  number={2},
  pages={166--183},
  year={1999},
  publisher={Sage Publications Sage CA: Thousand Oaks, CA}
}

@article{scarpa2000numerical,
  title={Numerical and experimental uniaxial loading on in-plane auxetic honeycombs},
  author={Scarpa, F and Panayiotou, P and Tomlinson, G},
  journal={The Journal of Strain Analysis for Engineering Design},
  volume={35},
  number={5},
  pages={383--388},
  year={2000},
  publisher={SAGE Publications Sage UK: London, England}
}

@article{dirrenberger2013effective,
  title={Effective elastic properties of auxetic microstructures: anisotropy and structural applications},
  author={Dirrenberger, Justin and Forest, Samuel and Jeulin, Dominique},
  journal={International Journal of Mechanics and Materials in Design},
  volume={9},
  pages={21--33},
  year={2013},
  publisher={Springer}
}

@article{yang2015mechanical,
  title={Mechanical properties of 3D re-entrant honeycomb auxetic structures realized via additive manufacturing},
  author={Yang, Li and Harrysson, Ola and West, Harvey and Cormier, Denis},
  journal={International Journal of Solids and Structures},
  volume={69},
  pages={475--490},
  year={2015},
  publisher={Elsevier}
}

@article{dos2012equivalent,
  title={Equivalent mechanical properties of auxetic lattices from discrete homogenization},
  author={Dos Reis, Francisco and Ganghoffer, Jean-Fran{\c{c}}ois},
  journal={Computational Materials Science},
  volume={51},
  number={1},
  pages={314--321},
  year={2012},
  publisher={Elsevier}
}

@article{kochmann2013homogenized,
  title={Homogenized mechanical properties of auxetic composite materials in finite-strain elasticity},
  author={Kochmann, Dennis M and Venturini, Gabriela N},
  journal={Smart materials and structures},
  volume={22},
  number={8},
  pages={084004},
  year={2013},
  publisher={IOP Publishing}
}

@article{slann2015cellular,
  title={Cellular plates with auxetic rectangular perforations},
  author={Slann, Alex and White, William and Scarpa, Fabrizio and Boba, Katarzyna and Farrow, Ian},
  journal={physica status solidi (b)},
  volume={252},
  number={7},
  pages={1533--1539},
  year={2015},
  publisher={Wiley Online Library}
}

@article{grima2016auxetic,
  title={Auxetic perforated mechanical metamaterials with randomly oriented cuts},
  author={Grima, Joseph N and Mizzi, Luke and Azzopardi, Keith M and Gatt, Ruben and others},
  journal={Adv. Mater},
  volume={28},
  number={2},
  pages={385--389},
  year={2016}
}

@article{ben2023gam,
  title={Gam: general auxetic metamaterial with tunable 3d auxetic behavior using the same unit cell boundary connectivity},
  author={Ben-Yelun, Ismael and G{\'o}mez-Carano, Guillermo and San Mill{\'a}n, Francisco J and Sanz, Miguel {\'A}ngel and Mont{\'a}ns, Francisco Javier and Saucedo-Mora, Luis},
  journal={Materials},
  volume={16},
  number={9},
  pages={3473},
  year={2023},
  publisher={MDPI}
}

@article{du2023auxetic,
  title={Auxetic kirigami metamaterials upon large stretching},
  author={Du, Chen and Wang, Yiqiang and Kang, Zhan},
  journal={ACS applied materials \& interfaces},
  volume={15},
  number={15},
  pages={19190--19198},
  year={2023},
  publisher={ACS Publications}
}

@article{chang2022machine,
  title={Machine learning-based inverse design of auxetic metamaterial with zero Poisson's ratio},
  author={Chang, Yafeng and Wang, Hui and Dong, Qinxi},
  journal={Materials Today Communications},
  volume={30},
  pages={103186},
  year={2022},
  publisher={Elsevier}
}

@article{liu2023high,
  title={High-efficient and reversible intelligent design for perforated auxetic metamaterials with peanut-shaped pores},
  author={Liu, Hongyuan and Hou, Feng and Li, Ang and Lei, Yongpeng and Wang, Hui},
  journal={International Journal of Mechanics and Materials in Design},
  volume={19},
  number={3},
  pages={553--566},
  year={2023},
  publisher={Springer}
}

@article{vyavahare2023fdm,
  title={FDM manufactured auxetic structures: an investigation of mechanical properties using machine learning techniques},
  author={Vyavahare, Swapnil and Teraiya, Soham and Kumar, Shailendra},
  journal={International Journal of Solids and Structures},
  volume={265},
  pages={112126},
  year={2023},
  publisher={Elsevier}
}

@article{tajalsir2022numerical,
  title={Numerical and random forest modelling of the impact response of hierarchical auxetic structures},
  author={Tajalsir, Ahmed Haytham and Mustapha, KB and Ibn-Mohammed, Taofeeq},
  journal={Materials Today Communications},
  volume={31},
  pages={103797},
  year={2022},
  publisher={Elsevier}
}

@article{wang2021novel,
  title={Novel planar auxetic metamaterial perforated with orthogonally aligned oval-shaped holes and machine learning solutions},
  author={Wang, Hui and Xiao, Si-Hang and Zhang, Chong},
  journal={Advanced Engineering Materials},
  volume={23},
  number={7},
  pages={2100102},
  year={2021},
  publisher={Wiley Online Library}
}

@article{liao2022deep,
  title={Deep-learning-based isogeometric inverse design for tetra-chiral auxetics},
  author={Liao, Zhongyuan and Wang, Yingjun and Gao, Liang and Wang, Zhen-Pei},
  journal={Composite Structures},
  volume={280},
  pages={114808},
  year={2022},
  publisher={Elsevier}
}

@article{lyngdoh2022elucidating,
  title={Elucidating the auxetic behavior of cementitious cellular composites using finite element analysis and interpretable machine learning},
  author={Lyngdoh, Gideon A and Kelter, Nora-Kristin and Doner, Sami and Krishnan, NM Anoop and Das, Sumanta},
  journal={Materials \& Design},
  volume={213},
  pages={110341},
  year={2022},
  publisher={Elsevier}
}

@article{zhang2024critical,
  title={A critical review on the application of machine learning in supporting auxetic metamaterial design},
  author={Zhang, Chonghui and Zhao, Yaoyao Fiona},
  journal={Journal of Physics: Materials},
  year={2024},
  publisher={IOP Publishing}
}

@article{moulinec1994fast,
  title={A fast numerical method for computing the linear and nonlinear mechanical properties of composites},
  author={Moulinec, Hervé and Suquet, Pierre},
  journal={Comptes Rendus de l'Acad{\'e}mie des sciences. S{\'e}rie II. M{\'e}canique, physique, chimie, astronomie},
  year={1994}
}

@article{moulinec1998numerical,
  title={A numerical method for computing the overall response of nonlinear composites with complex microstructure},
  author={Moulinec, Hervé and Suquet, Pierre},
  journal={Computer methods in applied mechanics and engineering},
  volume={157},
  number={1-2},
  pages={69--94},
  year={1998},
  publisher={Elsevier}
}

@article{schneider2021review,
  title={A review of nonlinear FFT-based computational homogenization methods},
  author={Schneider, Matti},
  journal={Acta Mechanica},
  volume={232},
  number={6},
  pages={2051--2100},
  year={2021},
  publisher={Springer}
}

@article{lucarini2021fft,
  title={FFT based approaches in micromechanics: fundamentals, methods and applications},
  author={Lucarini, Sergio and Upadhyay, Manas V and Segurado, Javier},
  journal={Modelling and Simulation in Materials Science and Engineering},
  volume={30},
  number={2},
  pages={023002},
  year={2021},
  publisher={IOP Publishing}
}

@article{danesh2023challenges,
  title={Challenges in two-scale computational homogenization of mechanical metamaterials},
  author={Danesh, Hooman and Brepols, Tim and Reese, Stefanie},
  journal={PAMM},
  volume={23},
  number={1},
  pages={e202200139},
  year={2023},
  publisher={Wiley Online Library}
}

@article{vondvrejc2014fft,
  title={An FFT-based Galerkin method for homogenization of periodic media},
  author={Vond{\v{r}}ejc, Jaroslav and Zeman, Jan and Marek, Ivo},
  journal={Computers \& Mathematics with Applications},
  volume={68},
  number={3},
  pages={156--173},
  year={2014},
  publisher={Elsevier}
}

@article{vondvrejc2015guaranteed,
  title={Guaranteed upper--lower bounds on homogenized properties by FFT-based Galerkin method},
  author={Vond{\v{r}}ejc, Jaroslav and Zeman, Jan and Marek, Ivo},
  journal={Computer Methods in Applied Mechanics and Engineering},
  volume={297},
  pages={258--291},
  year={2015},
  publisher={Elsevier}
}

@article{lucarini2022adaptation,
  title={Adaptation and validation of FFT methods for homogenization of lattice based materials},
  author={Lucarini, Sergio and Cobian, Lucia and Voitus, A and Segurado, J},
  journal={Computer Methods in Applied Mechanics and Engineering},
  volume={388},
  pages={114223},
  year={2022},
  publisher={Elsevier}
}

@article{willot2015fourier,
  title={Fourier-based schemes for computing the mechanical response of composites with accurate local fields},
  author={Willot, Fran{\c{c}}ois},
  journal={Comptes Rendus M{\'e}canique},
  volume={343},
  number={3},
  pages={232--245},
  year={2015},
  publisher={Elsevier}
}

@article{gierden2022review,
  title={A review of FE-FFT-based two-scale methods for computational modeling of microstructure evolution and macroscopic material behavior},
  author={Gierden, Christian and Kochmann, Julian and Waimann, Johanna and Svendsen, Bob and Reese, Stefanie},
  journal={Archives of Computational Methods in Engineering},
  volume={29},
  number={6},
  pages={4115--4135},
  year={2022},
  publisher={Springer}
}

@article{idrissi2022multiparametric,
  title={Multiparametric modeling of composite materials based on non-intrusive PGD informed by multiscale analyses: Application for real-time stiffness prediction of woven composites},
  author={Idrissi, M El Fallaki and Praud, Francis and Champaney, Victor and Chinesta, Francisco and Meraghni, Fodil},
  journal={Composite Structures},
  volume={302},
  pages={116228},
  year={2022},
  publisher={Elsevier}
}

@article{gokuzum2018algorithmically,
  title={An algorithmically consistent macroscopic tangent operator for FFT-based computational homogenization},
  author={G{\"o}k{\"u}z{\"u}m, Felix Selim and Keip, Marc-Andr{\'e}},
  journal={International Journal for Numerical Methods in Engineering},
  volume={113},
  number={4},
  pages={581--600},
  year={2018},
  publisher={Wiley Online Library}
}

@article{rambausek2019two,
  title={A two-scale FE-FFT approach to nonlinear magneto-elasticity},
  author={Rambausek, Matthias and G{\"o}k{\"u}z{\"u}m, Felix Selim and Nguyen, Lu Trong Khiem and Keip, Marc-Andr{\'e}},
  journal={International Journal for Numerical Methods in Engineering},
  volume={117},
  number={11},
  pages={1117--1142},
  year={2019},
  publisher={Wiley Online Library}
}

@article{minh2020surrogate,
  title={A surrogate model for computational homogenization of elastostatics at finite strain using high-dimensional model representation-based neural network},
  author={Minh Nguyen-Thanh, Vien and Trong Khiem Nguyen, Lu and Rabczuk, Timon and Zhuang, Xiaoying},
  journal={International Journal for Numerical Methods in Engineering},
  volume={121},
  number={21},
  pages={4811--4842},
  year={2020},
  publisher={Wiley Online Library}
}

@article{reutskiy2012method,
  title={The method of approximate fundamental solutions (MAFS) for elliptic equations of general type with variable coefficients},
  author={Reutskiy, S Yu},
  journal={Engineering analysis with boundary elements},
  volume={36},
  number={6},
  pages={985--992},
  year={2012},
  publisher={Elsevier}
}

@article{PLAprop,
  title={Numerical and experimental investigation of FDM fabricated re-entrant auxetic structures of ABS and PLA materials under compressive loading},
  author={Vyavahare, Swapnil and Kumar, Shailendra},
  journal={Rapid Prototyping Journal},
  volume={27},
  number={2},
  pages={223--244},
  year={2021},
  publisher={Emerald Publishing Limited}
}

@article{breiman2001random,
  title={Random forests},
  author={Breiman, Leo},
  journal={Machine learning},
  volume={45},
  pages={5--32},
  year={2001},
  publisher={Springer}
}

\end{document}